\documentclass[fleqn]{2023SCGE}
\usepackage{epstopdf}
\usepackage{graphicx}
\usepackage{epsfig}

\usepackage{mwe,tikz,ulem}
\usepackage[percent]{overpic}
\usepackage{graphicx}
\usepackage{xspace}
\usepackage[numbers]{natbib}
\usepackage[utf8x]{inputenc}

\usepackage[gen]{eurosym}
\usepackage{microtype}
\usepackage{color}
\usepackage{txfonts}
\usepackage{makeidx}
\usepackage[columns=1]{idxlayout}

\begin{document}
\ensubject{subject}

\ArticleType{Article}
\SpecialTopic{SPECIAL TOPIC: }
\Year{ }
\Month{ }
\Vol{ }
\No{ }
\DOI{ }
\ArtNo{ }
\ReceiveDate{April 2025 (Version 1) }


\title{The enhanced X-ray Timing and Polarimetry mission - eXTP for launch in 2030}{the eXTP mission}

\author[1,2]{\\Shuang-Nan Zhang}{zhangsn@ihep.ac.cn}
\author[3,1,30]{Andrea Santangelo}{}
\author[1]{Yupeng Xu}{}
\author[1]{Hua Feng}{}
\author[1]{Fangjun Lu}{}
\author[1]{\\Yong Chen}{}
\author[1]{Mingyu Ge}{}
\author[4]{Kirpal Nandra}{}
\author[5]{Xin Wu}{}
\author[6,7]{Marco Feroci}{}
% ...

\AuthorMark{Zhang S.N., Santangelo A., Xu Y. }


\AuthorCitation{Zhang S.N., Santangelo A., Xu Y., et al.}


\address[1]{State Key Laboratory for Particle Astrophysics, Institute of High Energy Physics, Beijing 100049, China}
\address[2]{University of Chinese Academy of Sciences, Beijing 100049, China}
\address[3]{Institut f\"{u}r Astronomie und Astrophysik, Eberhard Karls Universit\"{a}t, 72076 T\"{u}bingen, Germany}
\address[4]{Max Planck Institute for Extraterrestrial Physics, Giessenbachstr. 1, 85748 Garching, Germany}
\address[5]{Department of Nuclear and Particle Physics, University of Geneva, CH-1211 Geneva, Switzerland}
\address[6]{INAF -- Istituto di Astrofisica e Planetologia Spaziali, Via Fosso del Cavaliere 100, I-00133 Roma, Italy} 
\address[7]{INFN -- Roma Tor Vergata, Via della Ricerca Scientifica 1, I-00133 Roma, Italy} 
% ...
\address[30]{Center for Astronomy and Astrophysics, Center for Field Theory and Particle Physics, and \\ Department of Physics, Fudan University, Shanghai 200438, China}
% ...
\abstract{In this paper we present the current status of the enhanced X-ray Timing and Polarimetry mission, which has been fully approved for launch in 2030. eXTP is a space science mission designed to study fundamental physics under extreme conditions of matter density, gravity, and magnetism. The mission aims at determining the equation of state of matter at supra-nuclear density, measuring the effects of quantum electro-dynamics, and understanding the dynamics of matter in strong-field gravity. In addition to investigating fundamental physics, the eXTP mission is poised to become a leading observatory for time-domain and multi-messenger astronomy in the 2030's, as well as providing observations of unprecedented quality on a variety of galactic and extragalactic objects. After briefly introducing the history and a summary of the scientific objectives of the eXTP mission, this paper presents a comprehensive overview of: 1) the cutting-edge technology, technical specifications, and anticipated performance of the mission's scientific instruments; 2) the full mission profile, encompassing spacecraft design, operational capabilities, and ground segment infrastructure.}



\maketitle


\begin{multicols}{2}
\section{Introduction}\label{sec:Introduction}

\textit{The enhanced X--ray Timing and Polarimetry mission} -- eXTP is a scientific space mission designed to study the state of matter under extreme conditions of density, gravity and magnetism \cite{Zhang2016,2019SCPMA..6229502Z,Santangelo2023}. 
Primary goals are the determination of the equation of state of matter at supra-nuclear density, the measurement of Quantum Electro-Dynamics (QED) effects in the radiation emerging from highly magnetized stars, and the study of matter dynamics in the strong-field regime of gravity. The matter inside neutron stars (NSs), the space-time close to Black Holes (BHs), and the extremely magnetized vacuum close to magnetars are among the uncharted territories of fundamental physics. The eXTP mission will revolutionize these areas of  fundamental research by high precision X-ray measurements of NSs across the magnetic field scale and BHs across the mass scale. 
In addition to investigating fundamental physics, eXTP will be a very powerful observatory for the time-domain and multi-messenger astronomy in the 2030's, as well as providing observations of unprecedented quality on a variety of galactic and extragalactic objects. The eXTP science case is described in the other five white papers included in this special issue \cite{WP2,WP3,WP4,WP5,WP6}. They address the four main scientific objectives of the mission and the observatory science. These are updated substantially from the white papers published in 2019 \cite{2019SCPMA..6229503W, 2019SCPMA..6229505S, 2019SCPMA..6229504D, 2019SCPMA..6229506I}, to account for the rapid developments of the field over the past several years and the new mission profile described in the present paper.

\begin{figure}[H]
\centering
\includegraphics[width=0.9\columnwidth, angle=0]{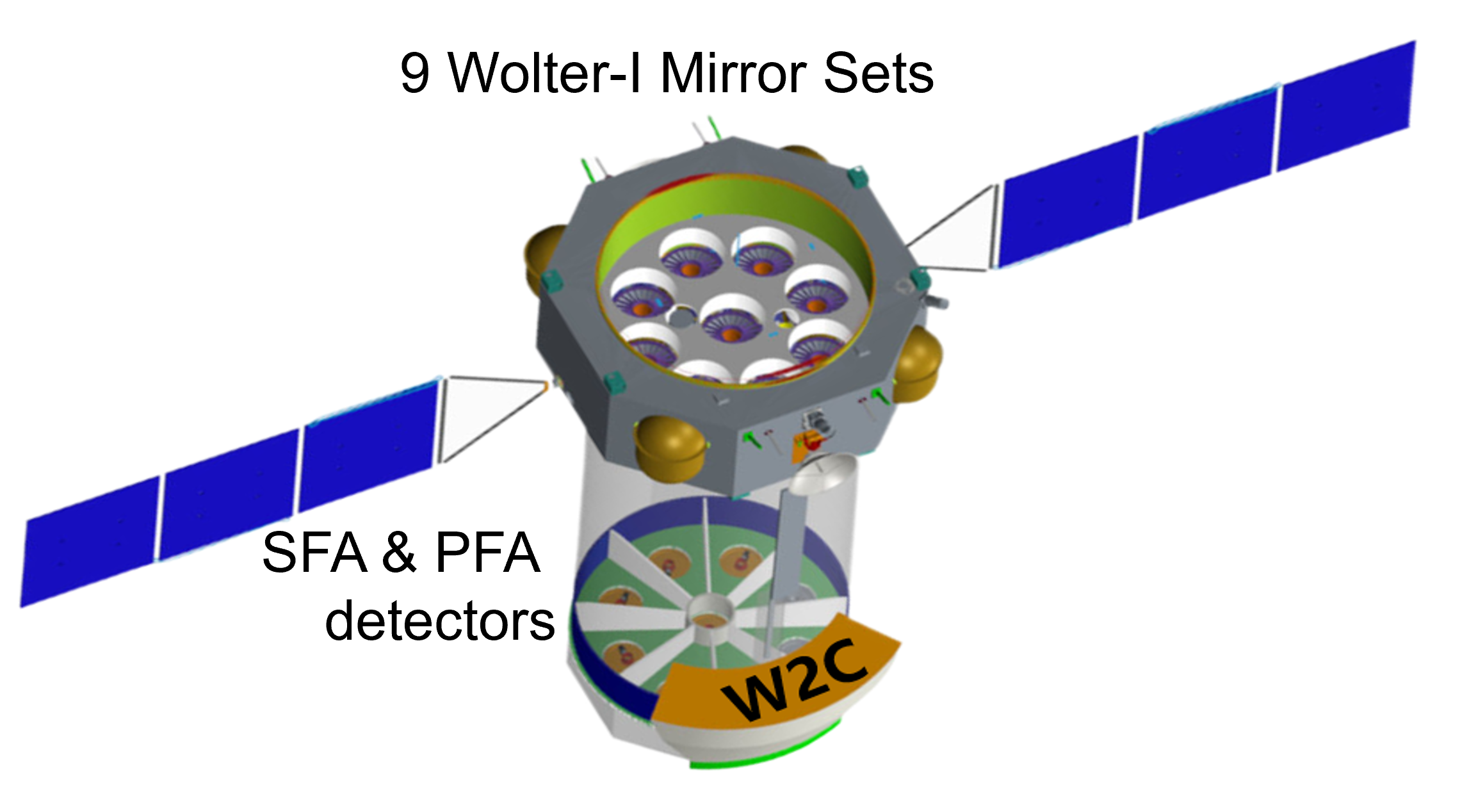}
\caption{Illustration of the eXTP satellite. The science payload consists of two main instruments and a secondary instrument: the focused SFA and PFA telescopes arrays, and the W2C to monitor a large fraction of the sky over a broad energy band.}
\label{Fig:eXTP_satellite}
\end{figure}

\begin{figure}[H]
\centering
\includegraphics[width=1.0\columnwidth]{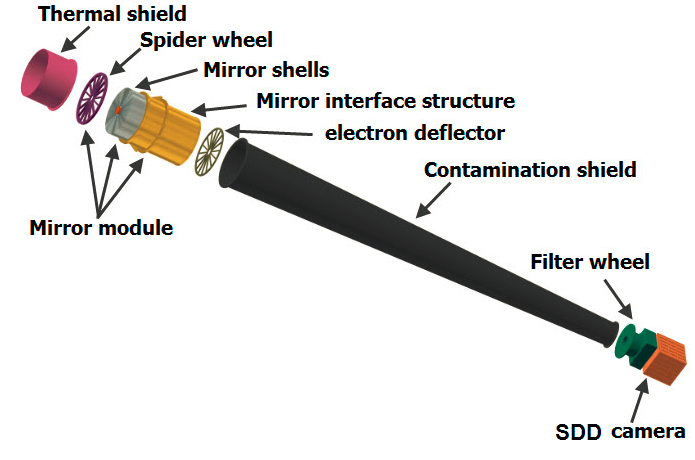}
\caption{The schematic structure of one of the SFA telescopes. (Taken from the previous eXTP mission paper \cite{2019SCPMA..6229502Z}.)}
\label{fig_SFA_ske}
\end{figure}

The eXTP mission is an enhanced version of the Chinese \textit{X-ray Timing and Polarimetry mission} (XTP) \cite{XTP}, which in 2011 was selected and funded for a Phase 0/A study as one of the background concept missions in the Strategic Priority Space Science Program of the Chinese Academy of Sciences (CAS). Also in 2011, the \textit{Large Observatory for Timing } (LOFT) mission concept \cite{Feroci2012, Feroci2014} was selected for an assessment study in the context of the ESA's Announcement of Opportunity for the third of the medium size missions (M3) foreseen in the framework of the Agency's Cosmic Vision program. The LOFT study was carried out in 2011-2014 by a consortium of European institutes. Eventually, the exoplanetary mission PLATO \cite{2014ExA....38..249R}, considered more well timed, was selected as the ESA Cosmic Vision M3 mission. Following this, in 2015, the European LOFT consortium and the Chinese team merged the concepts of the LOFT and XTP missions, thus starting the eXTP project. 

 The mission completed an extended phase A study in 2019 and a Phase B study in 2023, respectively. Due to the uncertainty in the availability of the expected European contributions, the mission was reconfigured near the end of 2023 to the current new baseline. The eXTP mission based on the new baseline was then approved by CAS in early 2024, approved at the national level in early 2025, and adopted in the middle of 2025 for launch in early 2030.

According to the adopted mission profile,  the Large Area Detector (LAD, a large array of collimated detectors made of silicon draft detectors (SDDs)) and the Wide Field Monitor (WFM, six modules of 1.5-D coded mask imager with SDDs as the pixelated detector planes) are not in the baseline. Both LAD and WFM are the main instruments of the LOFT mission \cite{Feroci2012, Feroci2014} and also the proposed European contributions to the previous baseline of the eXTP mission \cite{2019SCPMA..6229502Z, Santangelo2023}.
In addition, the number of telescopes of the Spectroscopy Focusing Array (SFA) and Polarimetry Focusing Array (PFA) has been reduced from 9 to 6 and from 4 to 3, respectively, to facilitate deployment using a smaller, more cost-effective launch vehicle; the focal plane camera of one of the six SFA telescopes is replaced by a pnCCD (Charge Coupled Device) camera, while the other five still use the original SDDs. The baseline orbit has been modified from a near-equatorial low-Earth orbit (LEO) to a high elliptical orbit (HEO) to enhance observational efficiency. However, a LEO configuration remains a viable alternative. A Wide Field and Wideband Camera (W2C) is considered as a secondary instrument to partially mitigate the loss of WFM capabilities.
The key performance parameters of the eXTP satellite's current design are summarized in Table \ref{tab:benchmark_instrument}, with comparative metrics from other flagship X-ray observatories listed for contextual benchmarking.

The eXTP consortium is led by the Institute of High Energy Physics (IHEP) of the CAS. In total, more than 200 scientists in more than 80 institutions from 17 countries are members of the eXTP instrument teams and science working groups.
In this paper, we present: (1) a summary of the scientific objectives of the eXTP mission in Section 2; (2) the technological implementation and expected performance characteristics of the scientific payload instruments in Section 3; and (3) the key operational elements and functional architecture of the mission in Section 4.


\section{Scientific objectives of the eXTP mission}\label{sec:sciencegoals}

\begin{table*}[htbp]
\caption{The key Instrument performances of the current design of the eXTP satellite and other main current/future missions for comparison. These parameters includes  effective area ($A_{\rm eff}$), energy range ($E$), timing resolution ($T$), polarimetric sensitivity (Minimum Detectable Polarization, $P_{\rm MDP}$), angular resolution ($\theta$), background level ($F_{\rm bkg}$) and pile-up limit ($PE$).}
\label{tab:benchmark_instrument}
\footnotesize
\begin{tabularx}{\textwidth}{lccccccc}
\toprule
Instrument & \textbf{$A_{\rm eff}$} & \textbf{$E$} & \textbf{$T$} & \textbf{$P_{\rm MDP}$} & \textbf{$\theta$} & \textbf{$F_{\rm bkg}$} &  $PE$ \\
& (cm$^{2}$) & (keV) & & (MDP) &  & (cts/s/cm$^2$/keV) & (cts/s)\\
\midrule
eXTP/SFA-T & 2750 @ 1.5 keV    & 0.5--10 & $10\,\mu$s & No & $\sim1'$ & $<1\times10^{-2}$ & No \\
&1670 @ 6 keV&&&&&& \\
eXTP/SFA-I & 550 @ 1.5 keV  & 0.5--10 & 50 ms (full)  & No & $\sim1'$ & $\sim10^{-2}$ & ~3 (FF)\\
& $330$ @ 6 keV&&   240 $\mu$s (timing)&&& & \\
eXTP/PFA & 250 @ 3 keV & 2--8 & $10\,\mu$s & 2\% (100 ks) & $\sim30''$ & $<10^{-2}$  & No\\
eXTP/W2C & 160 @ 60 keV & 30--600 & $\leq1$ ms & No & $<1^\circ$ & $\sim10^{-2}$  & No\\
\midrule
RXTE/PCA\cite{1996SPIE.2808...59J} & 6500 @ 6 keV  & 2--60 & $1\,\mu$s & No & $\sim1^\circ$ & $\sim10^{-1}$  & No \\
&  $\sim3000$ @ 3 keV&&&&&& \\
Chandra (ACIS-I)\cite{2002PASP..114....1W}& $\sim600$ @ 1.5 keV & 0.2--10 & 3.2s & No & $\sim0.5''$ & $\sim10^{-3}$ & 0.03 \\
XMM-Newton (EPIC-PN)\cite{Chambure}& $\sim1500$ @ 1.5 keV & 0.15--12 & 30 $\mu$s (timing) & No & $\sim15''$ & $\sim10^{-3}$ & 8 \\
&  $\sim900$ @ 6 keV &&&&&& \\
NICER\cite{NICER} & $\sim1900$ @ 1.5 keV  & 0.2--12 & $<0.3\,\mu$s & No & $\sim6'$ (non-imaging) & $\sim10^{-2}$  & No\\
&  $\sim600$ @ 6 keV &&&&&& \\
IXPE\cite{IXPE}& $\sim60$ @ 3 keV & 2--8 & $\sim10\,\mu$s & $\sim3\%$ ( 100 ks) & $\sim30''$ & $<10^{-2}$  & No \\
EP-FXT\cite{Yuan}& $\sim600$ @ 1.25 keV & 0.5--10 & $\sim44\,\mu$s & - & $\sim30''$ & $<10^{-2}$  & ~5 (FF)\\
NewAthena/X-IFU\cite{NewAthena} & $\sim5200$ @ 1 keV & 0.2--12 & 10 $\mu$s & No & $\sim9''$ & $\sim10^{-3}$ & -\\
NewAthena/WFI\cite{NewAthena} & $\sim8600$ @ 1 keV & 0.2--15 & 80 $\mu$s (fast)  & No & $\sim9''$ & $\sim10^{-3}$ &-  \\
&&& 160 $\mu$s (normal)&&&& \\
\bottomrule
\end{tabularx}
\end{table*}



The scientific objectives of eXTP can be divided into five themes: 1) to investigate the equation of state (EoS) of ultra-dense matter in NSs, presented in detail in White Paper 2 (WP2, \cite{WP2});
2) to understand black holes by probing strong gravity regions through synergistic X-ray spectral, timing, and polarimetric observations, presented in details in White Paper 3 (WP3, \cite{WP3}); 3) to resolve some long-standing questions about QED effects and radiative processes in the strong magnetic fields, presented in details in White Paper 4 (WP4, \cite{WP4}); 4) to play a key role in the era of time-domain and multi-messenger astronomy, presented in details in White Paper 5 (WP5, \cite{WP5}); and 5) to provide observations of unprecedented quality on a variety of galactic and extragalactic objects, as a very powerful observatory for astrophysics, presented in details in White Paper 6 (WP6, \cite{WP6}). Scientific goals in themes 1-4 are the core objectives of the current eXTP mission, expanded from the original themes 1-3 \cite{2019SCPMA..6229502Z}, which have driven the eXTP baseline mission design and its evolution.

WP2 focuses on how to determine the relation between the mass and radius of NSs with the eXTP observations, which are the key elements to studying EoS. Several methods of mass and radius measurements are proposed by different phenomena including: (1) mass and radius measurements by pulse profile modeling of rotation powered milli-second pulsars (MSPs); (2) mass and radius measurements by pulse profile modeling of accretion-powered MSPs together with the polarized information; (3) mass and radius constraints by thermonuclear burst cooling tails, accretion disk dynamics and cooling curve of MSPs; (4) the inner structure of NS revealed by glitches; and (5) EoS constraints with other phenomena and methods.

WP3 studies on how to understand black holes and the physics in strong gravity with eXTP observations. Leveraging its advanced instruments, eXTP will enable unprecedented precision in measuring black hole parameters and testing fundamental physics. The scientific objectives include: (1) spin and mass measurements using relativistic reflection spectroscopy (e.g., Fe $K_{\alpha}$ line profiles), thermal continuum fitting or polarization data; (2) measuring spins and masses of supermassive black holes (SMBHs) with quasi-periodic oscillations (QPOs), X-ray reverberation in super-Eddington accretion flows, and track orbital hotspots; (3) testing general relativity by analyzing possible deviations from Kerr metrics; 
 and (4) accretion physics, such as disk-corona dynamics and coronal geometry, with time-resolved spectral-timing techniques and polarization data.

WP4 discusses how to resolve some long-standing questions about QED effects and radiative processes in strong magnetic fields with eXTP observations. The primary science goals include: (1) testing QED predictions, such as vacuum birefringence, through polarization studies of magnetars and neutron stars with ultra-strong magnetic fields (\(10^{13}-10^{15}\) G); (2) probing magnetic field topology and surface emission mechanisms in magnetars and accreting X-ray pulsars (XRPs) via phase-resolved polarimetry and spectroscopy; (3) investigating accretion dynamics in strongly magnetized systems, including transitions between sub/super-critical accretion regimes and the geometry of accretion columns; (4) characterizing magnetar outbursts, glitches, and their connection to fast radio bursts (FRBs); and (5) exploring extreme accretion in pulsating ultraluminous X-ray sources (PULXs).

WP5 demonstrates the roles eXTP may play in the fast rising time-domain and multi-messenger astronomy through its advanced instruments. Key science topics include: (1) probing gravitational wave (GW) counterparts such as neutron star/black hole mergers and extreme mass-ratio inspirals (EMRIs) to constrain neutron star equations of state and black hole spins, studying SMBHs via X-ray timing and Fe $K_{\alpha}$ line diagnostics, and exploring gamma-ray burst (GRB) physics through polarization measurements to distinguish synchrotron versus photospheric emission mechanisms and test quantum gravity effects via X-ray polarimetry; (2) investigating magnetar bursts, FRBs, and long-period radio transients to unravel magnetic field dynamics and emission origins; (3) monitoring tidal disruption events (TDEs) to understand super-Eddington accretion and relativistic jet formation; and (4) observation of core-collapse supernovae, TeV-active galactic nuclei, and potential neutrino-associated transients.

WP6 showcases that eXTP, as a very powerful observatory for astrophysics, will provide observations of unprecedented quality on a variety of galactic and extragalactic objects. The observatory science objectives may include: characterizing stellar coronae and transient phenomena such as flares and coronal mass ejections, probing magnetic turbulence and particle acceleration in supernova remnants and pulsar wind nebulae, investigating accretion physics across diverse systems (e.g., Cataclysmic Variables (CVs), X-ray binaries (XRBs), Active Galactic Nuclei (AGNs)), exploring super-Eddington accretion in ultraluminous X-ray sources (ULXs), and advancing X-ray pulsar-based timing techniques.


\section{The scientific payload}\label{sec:2}

An illustration of the current design of the eXTP satellite is shown in Figure~\ref{Fig:eXTP_satellite}. 
The scientific payload of the mission consists of two main and possibly one secondary instruments: the SFA (Section \ref{sec:SFA}), the PFA (Section \ref{sec:PFA}), and the W2C (Section \ref{sec:W2C}).

\subsection{Spectroscopy Focusing Array - SFA}\label{sec:SFA}

The Spectroscopic Focusing Array (SFA) comprises six Wolter-I grazing-incidence X-ray telescopes designed for spectral, timing, and imaging observations in the 0.5 to 10\,keV energy range. Each telescope assembly includes
the mirror module, the electron deflector, the filter wheel, and the focal plane camera. 
Five of the focal plane cameras are equipped with SDDs, and their corresponding telescopes are designated SFA-T (where 'T' denotes 'Timing'), reflecting the SDDs' excellent timing resolution. A schematic of the SFA-T telescope design is presented in Figure~\ref{fig_SFA_ske}.
The sixth telescope features a pnCCD focal plane camera, analogous to those employed in eROSITA \cite{Predhel} and Einstein Probe \cite{Yuan}, providing enhanced point-source sensitivity (via reduced background) and superior imaging performance. This unit is designated SFA-I (where `I' signifies `Imaging').
The focal plane detector modules for both the SFA-T and SFA-I detectors will be developed and contributed to the eXTP project by the Max Planck Institute for Extraterrestrial Physics (MPE). The sensor chips are fabricated by the Semiconductor Laboratory of the Max Planck Society (MPG-HLL). 
The total effective area of SFA is expected to be larger than $\sim3300$\,cm$^2$ at 1.5\,keV and $\sim2000$\,cm$^2$ at 6\,keV, and the field of view (FoV) is 18\,arcmin. 
Each SFA-T telescope includes a 19-cell SDD array, whose energy resolution is better than 180\,eV at 6\,keV. The time resolution is 10\,$\mu$s, and the dead time is expected to be less than 5\% for a source flux of approximately 1\,Crab. The angular resolution must be less than 1\,arcmin (half-power diameter, HPD).
To meet the requirements, the working temperature of the mirror modules must be stable at 20 $\pm$ 2\,$^{\circ}$C. 
An electron deflector is installed at the bottom of the mirror module to reduce the particle background.


\subsubsection{Optics}

Since they are based on very similar concepts, in this section, we discuss the optics of both the SFA and the PFA together. Nine X-ray grazing-incidence Wolter I (parabola + hyperbola) optics modules will be implemented onboard eXTP; the mirror technique implemented is nickel replication, which has already been successfully used for high-throughput X-ray telescopes with good angular resolution, such as for the BeppoSAX \cite{Citterio}, XMM-Newton \cite{Chambure}, Swift \cite{Burrows}, eROSITA \cite{Predhel} and  ART-XC telescopes onboard the Spektr-Roentgen-Gamma Mission  \cite{2021A&A...650A..42P}, the IXPE \cite{Ramsey,IXPE} polarimetric mission, and the FXT telescope aboard the Einstein Probe mission \cite{Yuan}. The SFA includes six optic modules, while three optic modules are used for the PFA. The expected collecting area of each telescope is $A_{\rm optics}$ $\gtrapprox820$ cm$^2$ (SFA) and $\gtrapprox800$ cm$^2$ (PFA) at 3\,keV, and $A_{\rm optics}$ $\gtrapprox550$ cm$^2$ at 6\,keV. 
The requirement for the angular resolution of the PFA optics is $\le 30$ arcsec (HPD)@3\,keV, more demanding than the angular resolution of 1 arcmin (HPD)@1.5\,keV required for the SFA optics. For both types of telescope, the focal length reference value is 525\,cm. The maximum diameter of the mirror shells is about 510 mm, constrained by the envelope of less than 600\,mm in diameter for a single mirror. In Table \ref{tab1} we list the main parameters of the SFA and PFA optics. 



\begin{figure}[H]
\centering
  \includegraphics[width=0.48\textwidth]{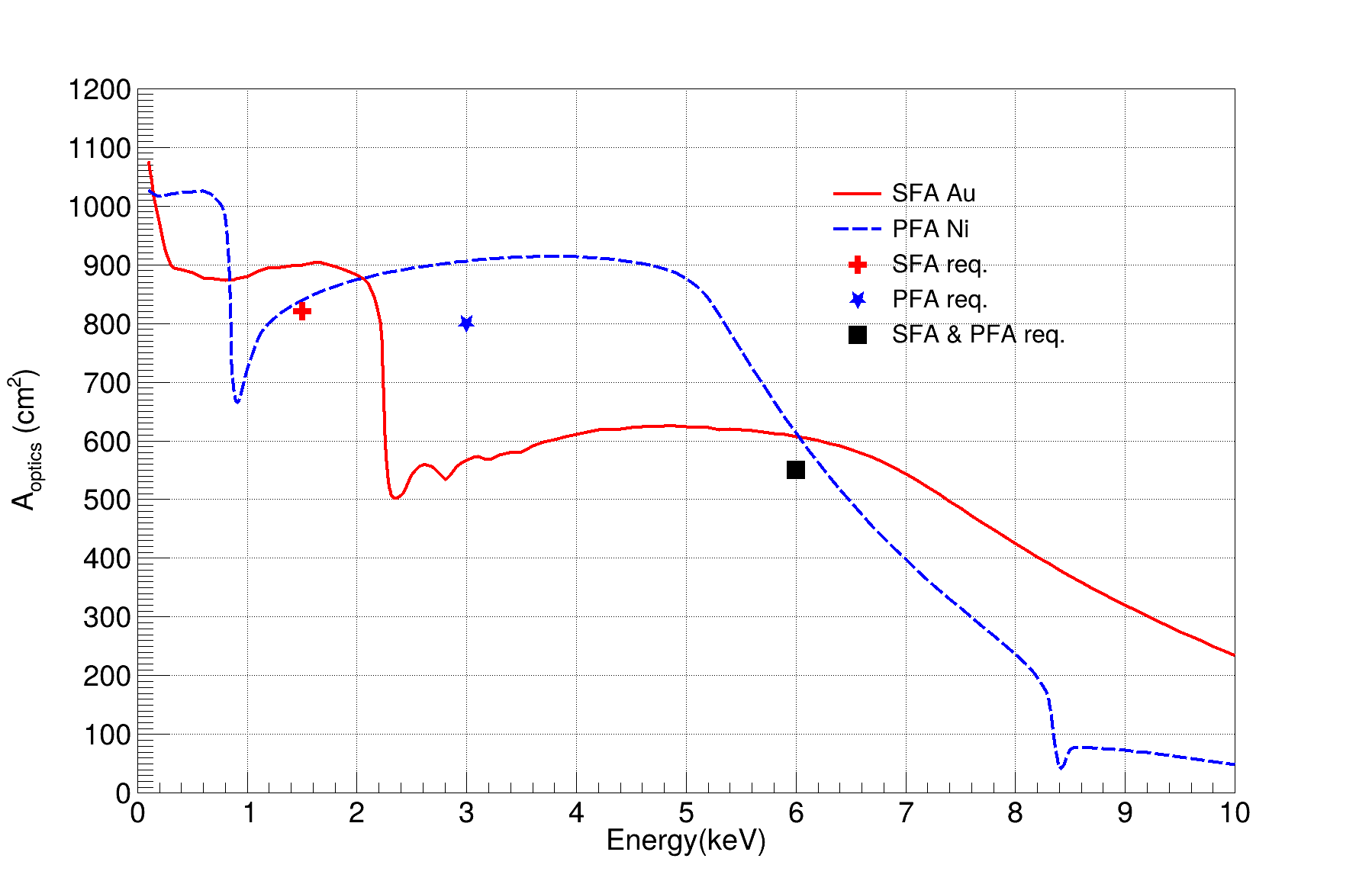}
\caption{On--axis collecting area $A_{\rm optics}$ expected for the eXTP X-ray optics based on Nickel electroforming (a single mirror module), with different coatings. For SFA, the gold film is selected with 100 nm thickness. For PFA, nickel is directly employed as the coating.}
  \label{fig:effectivearea}  
\end{figure}

\begin{figure}[H]
\centering
  \includegraphics[width=\columnwidth]{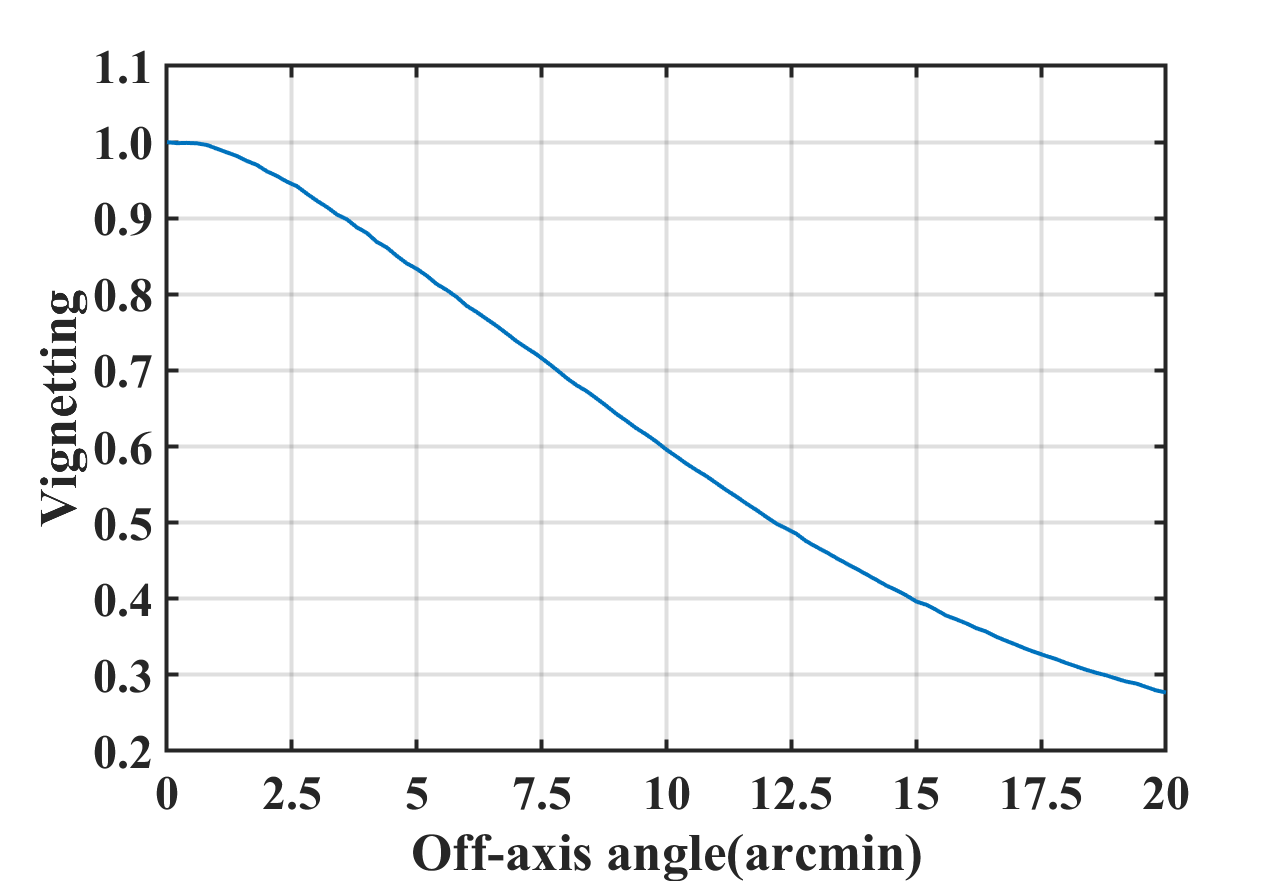}
\caption{Vignetting function at 1.5\,keV of the eXTP X-ray optics based on Nickel electroforming with Au coating.}
  \label{fig:vignetting}
\end{figure}

Figure \ref{fig:effectivearea} shows the expected collecting area $A_{\rm optics}$ of a mirror module with different coatings. For SFA, the gold film is selected with 100 nm thickness as the reflective material based on the following considerations: 1) the gold film exhibits stable reflectivity characteristics in the 0.2--10 keV energy range; 2) its thermal expansion coefficient closely matches that of nickel, facilitating mold release in the electroformed nickel gold-plating process; and 3) the gold-plating process offers high maturity and reliability. For PFA, nickel is directly employed as the reflective material to fulfill its requirements for effective area and quality factor. Figure \ref{fig:vignetting} shows the mirror vignetting function. 

\begin{table}[H] 
\begin{center}
\caption{The main parameters of the SFA and PFA baseline optics.}
\label{tab1}
\footnotesize
\begin{tabular}{l|l}
\bottomrule
\textbf{Parameters} & \textbf{For one telescope} \\
\hline
Focal length & 5.25\,m \\
Aperture & $\le$510\,mm (diameter) \\
Envelope & $\le$600\,mm (diameter) \\
$A_{\rm optics}$ of SFA & $\ge$820\,cm$^2$@1.5\,keV \\
$A_{\rm optics}$ of PFA & $\ge$800\,cm$^2$@3\,keV \\
$A_{\rm optics}$ of SFA \& PFA & $\ge$550\,cm$^2$@6\,keV \\
Energy range & 0.5--10\,keV \\
Field of view & $18^\prime \times 18^\prime$ \\
Angular resolution SFA & $\le 1^\prime$ (HPD)@1.5\,keV\\
Angular resolution PFA & $\le 30^{\prime\prime}$ (HPD)@3\,keV\\
\bottomrule
\end{tabular} 
\end{center}
\end{table}

\subsubsection{SFA-T Detectors}
The scientific objectives of the SFA require a detector with high count rate capability and time resolution, along with excellent spectroscopic performance. The energy resolution must be better than 180 eV (FWHM at 6 keV) until the end of the mission after 8 years of operation, and the time resolution must be less than 10\,$\mu$s. The main specifications for SFA-T (5 telescopes with SDD detectors) are listed in Table \ref{tab:SFA-T}.

The focal plane detector for each SFA-T telescope consists of an SDD sensor arranged in 19 hexagonal cells, with readout provided by custom ASICs, three of which are needed for each 19-cell SDD \cite{2024SPIE13093E..6QA}. 

This configuration meets the aforementioned requirements (see Fig.~\ref{fig:sdd_pnccd}). Compared with CCDs, SDDs allow a much faster readout. The detector geometry ensures that the vast majority of source photons are focused onto the inner 7 cells of the array, whereas the cells of the outermost ring are used to accurately determine the background. Thus, the side length of a hexagonal cell has been determined to be 3.2~mm given the expected angular resolution of the optics, a trade-off between the number of pixels and some sampling of the mirror’s response.. The sensor has a 450\,$\mu$m sensitive thickness, which delivers excellent quantum efficiency throughout the SFA energy range. The signal processing chain of each SDD cell includes a charge-sensitive preamplifier, a fast shaper, and a slow shaper with sample-and-hold circuit. A Field Programmable Gate Array (FPGA) based back-end electronics (BEE) module is implemented to control the front-end electronics (FEE) and read the digitized data from the analog-to-digital converters (ADCs). 

A filter wheel is located above the detector, with four positions corresponding to 4 operational modes: (1) \textit{Open filter}, used in ground tests and weak sources observations; (2) \textit{Calibration}. An Fe-55 radioactive source is used for in orbit calibration; (3) \textit{Optical blocking filter}. This filter, which consists of a 400\,nm polyimide and a 200\,nm aluminum film, is used to block visible and ultra-violet (UV) light. It also prevents contamination of the detector; (4) \textit{Closed filter}. A metal shutter is used to measure the internal background and prevents the detector from being damaged during launch and extreme solar events. To compensate for particle damage to the SDDs, especially due to protons, a temperature of $-55^{\circ}$ C is expected.
during the operation of the detectors. Temperature stability of $\pm$0.5\,$^{\circ}$C is also required. 

Please refer to \cite{2019SCPMA..6229502Z, Santangelo2023} for more details on SDDs and electronics.


\begin{table}[H]
\centering
\caption{SFA-T specifications (5 telescopes with SDD at focal plane), where both HPD and W90 (width of the point spread function enclosing 90\% reflected photons) are at 1.5 keV.}
\label{tab:SFA-T}
\footnotesize
\begin{tabular}{l|l}
\bottomrule
\hline
\textbf{Indicator} & \textbf{Requirement} \\ \hline
Total Effective Area & $\ge$2750 cm$^2$@1.5\,keV \\ 
(within $E\pm$0.5 keV) & $\ge$1670 cm$^2$@6\,keV \\ \hline
Energy Range & 0.5-10\,keV \\ \hline
Energy Resolution (FWHM) & $\le$180 eV@6\,keV \\ \hline
Field of View & $18^\prime$ (diameter) \\ \hline
SDD Pixel Size & 6.4 mm ($4.2^\prime$) \\ \hline
Angular Resolution & $\le 1^\prime$ (HPD), $3^\prime$ (W90) \\ \hline
Time Resolution & $\le 10 \,\mu$s \\ \hline
Absolute Timing Accuracy & $\le 2\,\mu$s \\ \hline
Dead Time & $\le 5$\%@1\,Crab \\ \hline
\end{tabular}
\end{table}


\subsubsection{SFA-I Detector}

\begin{figure}[H]
\centering
\includegraphics[width=0.8\columnwidth]{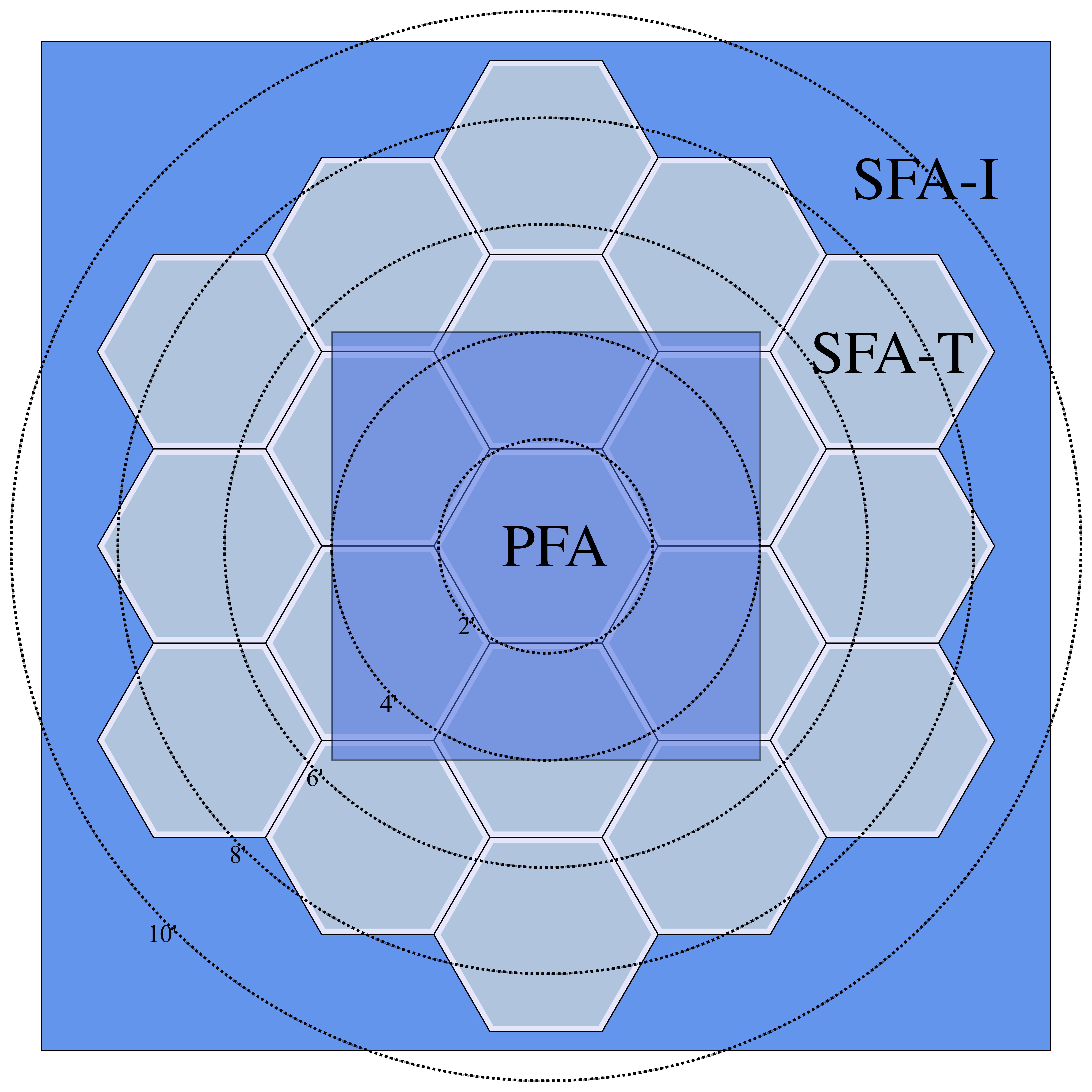}
\caption{Comparison between the FoVs of SDD (grey color, each of the 19 hexagon cells has a side length of 3.2\,mm and a sensitive thickness of 450\,$\mu$m), pnCCD (blue color) and GPD (blue color in the center) cameras.  Circles denote the angular scale in 2$^\prime$ increments from 2$^\prime$ to 10$^\prime$. The FoVs of PFA, SFA-T and SFA-I are thus 8$^\prime\times$8$^\prime$ (square), 18$^\prime$ in diameter (circle), and slightly larger than 18$^\prime\times$18$^\prime$ (square), respectively.  }
\label{fig:sdd_pnccd}
\end{figure}

The SFA-I pnCCD camera supports three distinct observation modes to address various scientific requirements. The key technical specifications include a 28.8×28.8\,mm$^2$ detector active area (5.25 m focal length) with 75×75\,$\mu$m$^2$ pixels (2.95$^{\prime\prime}$/pixel), where pileup constraints dictate mode selection for bright source observations.  The choice of a $\sim3^{\prime\prime}$ pixel size given the 1$^\prime$ HPD mirror module requirement is not an optimal solution. It is done so because we simply decided to replace the SDD focal plane camera with the existing pnCCD camera used for EP/FXT. However, we will select the SFA mirror module with the best angular resolution for SFA-I, to better match the $\sim3^{\prime\prime}$ pixel size of the pnCCD camera. The full frame mode serves wide-field surveys and multiobject spectroscopy, while the windowed mode facilitates time-resolved studies of medium-brightness sources. The timing mode specializes in millisecond variability analysis of bright transients. The main specifications for SFA-I (1 telescope with a pnCCD camera) are listed in Table \ref{tab:SFA-I}. The comparison between the FoVs of the SDD, pnCCD and gas pixel detector (GPD, for PFA) cameras is shown in Figure \ref{fig:sdd_pnccd}. The FoV of SFA-I (with pnCCD) is the largest (18 arcmin) and thus covers that of SFA-T and PFA, allowing SFA-I to provide simultaneous spectra (and images) of all sources in the FoVs of SFA-T and PFA.

\begin{figure}[H]
\centering
\includegraphics[width=1.0\columnwidth]{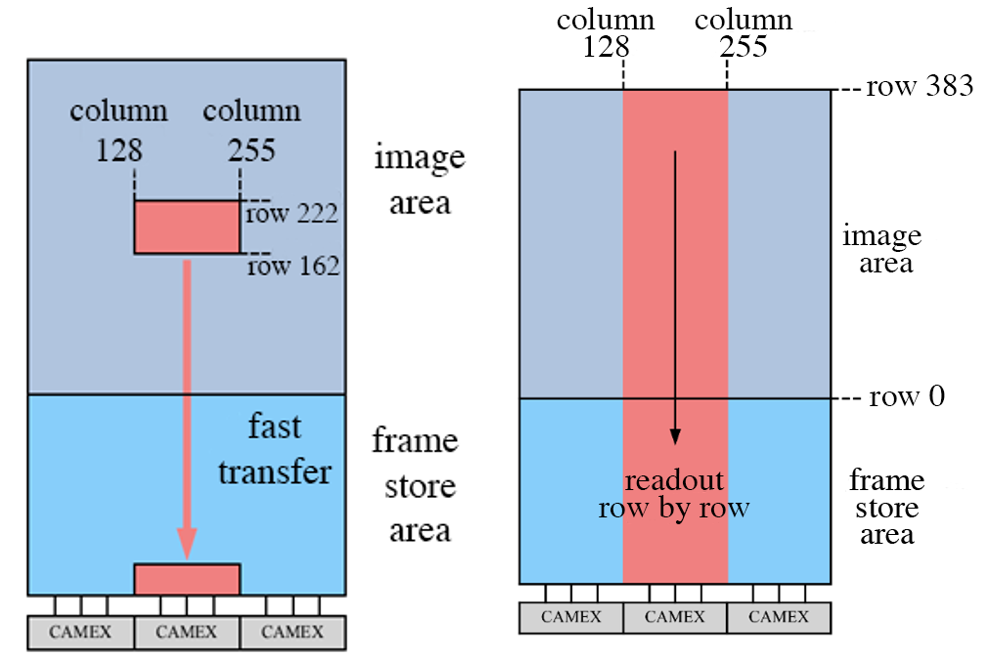}
\caption{Left: pnCCD windowed mode (61×128 pixels). Right: pnCCD timing mode (384×128 pixels fast readout).}
\label{fig:window}
\end{figure}

In full-frame mode (384×384 pixels) it provides an 18$^\prime$×18$^\prime$ field of view with 2.95$^{\prime\prime}$ per pixel resolution, simultaneously acquiring 0.5--10\,keV energy spectra while monitoring source variability at 50 ms time resolution and localizing targets with $<10^{\prime\prime}$ accuracy (90\% confidence). This mode supports observations up to 3 mCrab brightness under 10\% pile-up conditions. The windowed mode (61×128 pixels), as shown in Figure \ref{fig:window} (left), enhances the time resolution to $<$3 ms (2.2 ms cycle), optimized for medium brightness point sources with observational limits of 60--70\,mCrab. For bright targets, the timing mode (384×128 pixels fast readout), as shown in Figure \ref{fig:window} (right), enables 240 $\mu$s time resolution using 1D imaging (HPD 30$^{\prime\prime}$), accommodating sources up to 700 mCrab. Inheriting EP-FXT technology, SFA-I combines imaging, spectroscopy, and timing capabilities.

The pnCCD within the FoV of the telescopes simultaneously receives signals from X-ray photons and charged particles. However, it is usually difficult to distinguish between the signals of X-ray photons and charged particles with the same energy depositions. The frame’s storage section is also part of the pnCCD, but it is not exposed to the telescope's FoV, and therefore only receives charged particles but does not receive X-ray photons. After half of an exposure, the charge in the storage area due to energy depositions of charged particles is transferred out. At the end of the exposure, the charge from the pixels in the FoV is transferred into the storage area and then transferred out from there. Consequently, the charge first transferred out of the storage area can be used to determine the in-orbit particle background spectra and counting rates. The peripheral FoV regions, where no source is detected, are used for diffuse X-ray background measurements. Compared with SFA-T (with SDD arrays), while sharing identical energy resolution and effective area (from common 450 $\mu$m depletion layer detectors and focusing mirrors), SFA-I demonstrates distinct advantages and limitations.  The SDD array has 19 pixels and the outer ring of 12 pixels are used for estimating the background. However, unknown contaminating sources within the FoV cannot be identified and removed cleanly, resulting in systematic errors of the background modeling. The imaging capability of pnCCD and nearly the same energy response with SDD allow pnCCD to identify and characterize any contaminating sources within the FoV, which serves as an input for the background modeling of SFA-T with SDDs. 

\begin{figure}[H]
\centering
\includegraphics[width=1.0\columnwidth]{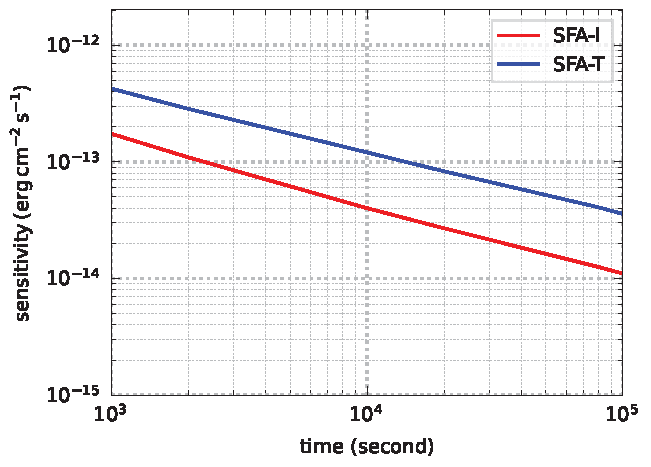}
\caption{Sensitivity (0.5-10 keV) comparison between one SFA-T telescope and one SFA-I.  The improved sensitivity of SFA-I is solely due to its smaller pixel size, so that lower background counts are recorded in the central pixels which receive most of the X-ray photons from the target point source. Note that the diffusion limits in some crowded fields due to the limited angular resolutions of the telescopes are not considered here, though negligible for SFA-I and not serious for SFA-T.}
\label{fig:sfa-i-sfa-t}
\end{figure}

The much cleaner background of SFA-I makes it 3 times more sensitive than of an SFA-T telescope for faint sources, as shown in Figure \ref{fig:sfa-i-sfa-t}; no comparison with the PFA sensitivity is made since the key requirement of PFA is its minimum detectable polarization degree. In WP2\cite{WP2}, some simulations are made to show how SFA-I may improve the estimate of the mass and radius of a NS by providing the simultaneous spectrum of a contaminating source in the FoV of SFA-T.
However, SFA-I exhibits a worse timing resolution compared to SDD and faces saturation challenges when observing bright sources such as BH and NS X-ray binary outbursts, as well as X-ray bursts of neutron stars and the bright early X-ray afterglows of gamma-ray bursts.

\begin{table}[H]
\centering
\caption{SFA-I specifications (1 telescope with pnCCD camera at focal plane), where both HPD and W90 are at 1.5 keV.}
\label{tab:SFA-I}
\footnotesize
\begin{tabular}{l|l}
\bottomrule
\hline
\textbf{Indicator} & \textbf{Requirement} \\ \hline
Total Effective Area & $\ge$550 cm$^2$@1.5\,keV \\ 
(within $E\pm$0.5 keV) & $\ge$330 cm$^2$@6\,keV \\ \hline
Energy Range & 0.5-10 keV \\ \hline
Energy Resolution (FWHM) & $\le$180\,eV@1.5\,keV \\ \hline
Field of View & $18^\prime \times 18^\prime$ \\ \hline
Angular Resolution & $\le 1^\prime$ (HPD), $3^\prime$ (W90) \\ \hline
Time Resolution & $\le$50 ms (full frame) \\ 
 & $\le$240  $\mu$s (timing mode) \\ 
 & $\le$3 ms (windowed mode) \\ \hline
\end{tabular}
\end{table}

%

\begin{figure}[H]
\centering
\includegraphics[width=1.\columnwidth]{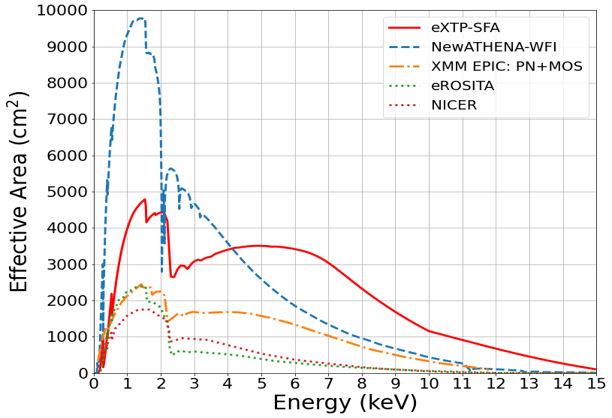}
\caption{The total effective area of the SFA (SFA-T+SFA-I), taking into account the optics and detector efficiency, in comparison with that of the WFI of the NewAthena, XMM-Newton and NICER missions.}
\label{fig_sfa_eff_area_new}
\end{figure}

The total effective area of the SFA (SFA-T+SFA-I), taking into account the detector efficiency and filters, is shown in Fig.~\ref{fig_sfa_eff_area_new}, compared to that of the Wide Field Imager (WFI) of the NewAthena, XMM-Newton and NICER missions \cite{NewAthena, Chambure, NICER}.

\subsection{Polarimetry Focusing Array -- PFA}
\label{sec:PFA}

The PFA consists of three identical telescopes optimized for X-ray imaging polarimetry, sensitive in the energy range of 2--8 keV. In synergy with the SFA, the PFA offers spatial, energy, and/or temporal resolved X-ray polarimetry at high sensitivity. 
The main PFA specifications are listed in Table \ref{tab:PFA}.

\begin{table}[H]
\centering
\caption{PFA specifications  (3 telescopes with GPD at focal plane)}
\footnotesize
\label{tab:PFA}
\begin{tabular}{l|l}
\bottomrule
\hline
\textbf{Indicator} & \textbf{Requirement} \\ \hline
Total Effective Area & $\ge$ 180 cm$^2$ @ 3 keV \\ \hline
Energy Range & 2 -- 8 keV \\ \hline
Energy Resolution (FWHM) & $\le$ 1.8 keV @ 6 keV \\ \hline
Field of View & $9.8^\prime \times 9.8^\prime$ \\ \hline
Angular Resolution & HPD $\le 30^{\prime\prime}$ @ 3 keV\\ \hline
Time Resolution & $\le $10 $\mu$s \\ \hline
Dead Time & $\le$ 10\% @ 1 Crab \\ \hline
Polarization Modulation Factor & $\ge$ 50\% @ 6 keV \\ \hline
Minimum Detectable Polarization & $\le$ 3\% (1 mCrab, 10$^6$ s)\\ \hline
\end{tabular}
\end{table}

\begin{figure}[H]
\centering
\includegraphics[width=\columnwidth]{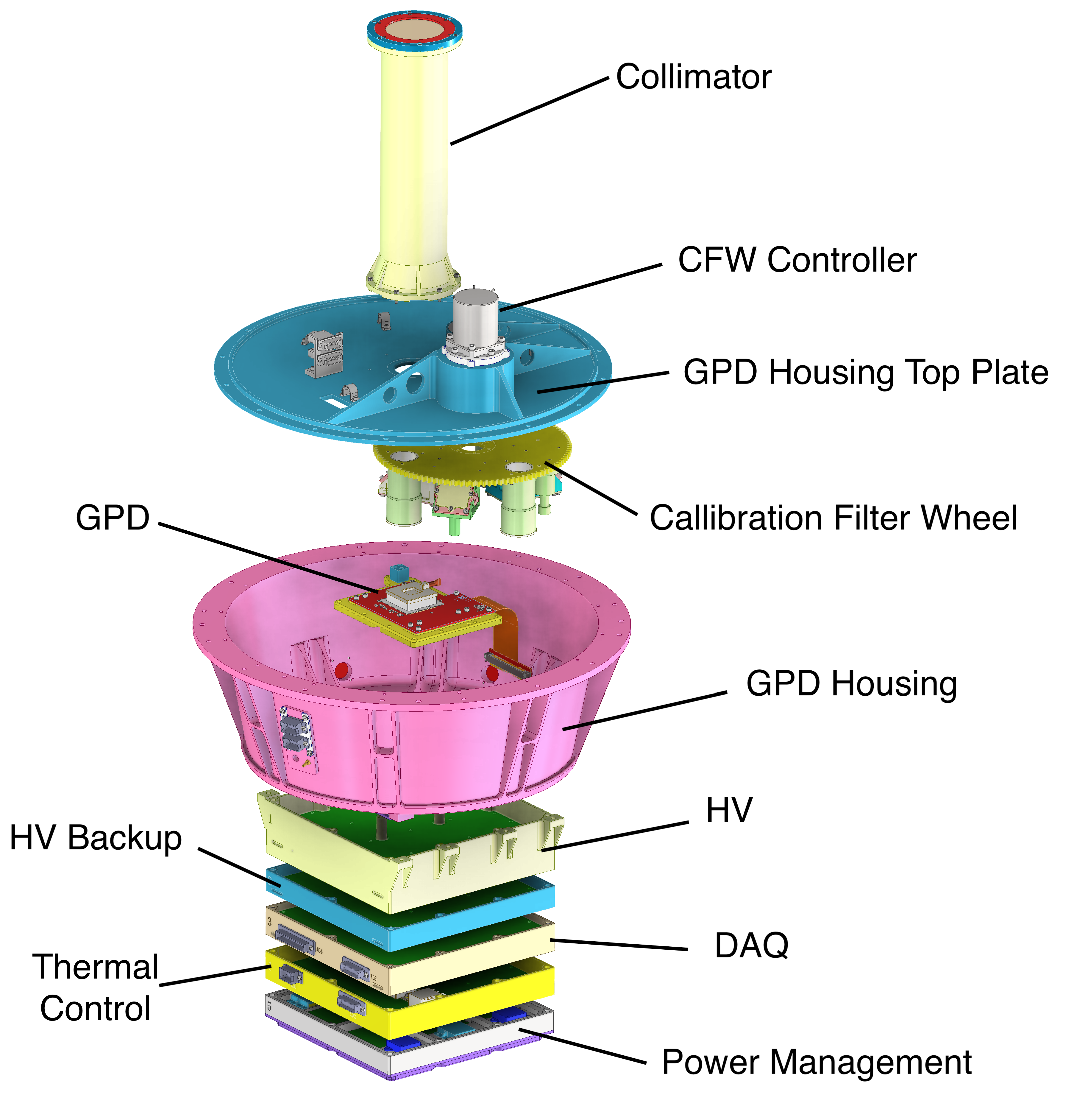}
\caption{An exploded view of the PFA detector unit.}
\label{fig:pfa_du}
\end{figure}

The PFA optics are described in Section~\ref{sec:SFA}. Figure~\ref{fig:pfa_du} provides an exploded view of the PFA detector unit (DU), which consists of three key components: the calibration and filter wheel in the top, the focal plane detector in the middle, and the back-end electronics at the bottom.

\subsubsection{Calibration and filter wheel}

\begin{figure}[H]
\centering
\includegraphics[width=0.9\columnwidth]{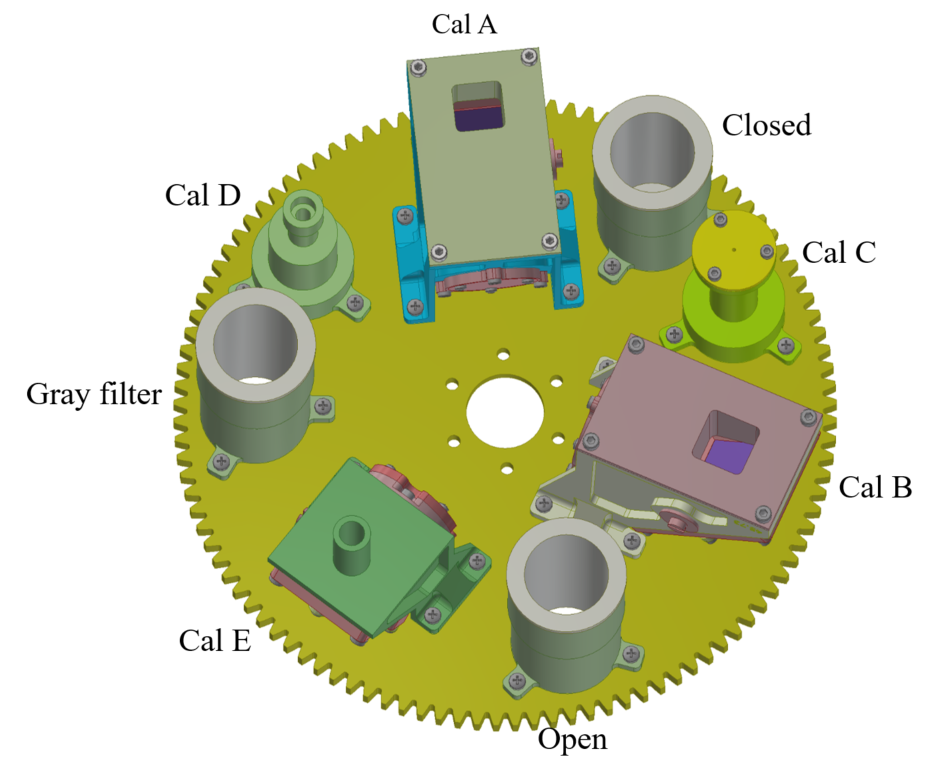}
\caption{A schematic drawing of the filter and calibration wheel.}
\label{fig:pfa_cfw}
\end{figure}

The calibration and filter wheel (CFW), shown in Figure~\ref{fig:pfa_cfw}, is to house five calibration sources and a set of filters (gray, open, and closed).  

The calibration sources will be used to calibrate or monitor the performance of the focal plane detector throughout the mission lifetime, including gain, energy resolution, modulation factor, and spurious modulation. Each contains a $^{55}$Fe radioactive isotope, which has a half-life of 2.7 years. Cal A and Cal B generate polarized X-rays via Bragg diffraction using different crystals. These two sources are designed to irradiate the entire sensitive area of the GPD and are mainly used to monitor or calibrate modulation factors. Cal A provides two energy lines at 3~keV and 5.9~keV, while Cal B offers an additional line at 4.5~keV. Cal C, Cal D, and Cal E are unpolarized sources. Cal C is a collimated 5.9~keV source, with a beam diameter of less than 5~mm, to monitor the gain, energy resolution, and spurious modulation in the central region of the GPD. In contrast, Cal D and Cal E are uncollimated and emit two lines at 5.9~keV and 1.7~keV, enabling performance monitoring across the entire sensitive area. Table~\ref{tab:pfa_calib} summarizes the characteristics of these calibration sources.

\begin{table}[H]
\centering
\caption{PFA in-orbit calibration sources}
\footnotesize
\label{tab:pfa_calib}
\begin{tabular}{l|l|l}
\bottomrule
\hline
\textbf{Name} & \textbf{Description} & \textbf{Rate} \\ \hline
Cal A & Polarized, 15 $\times$ 15 mm$^2$ & $>$ 5.2 c/s @ 3 keV \\
      &  & $>$ 1.6 c/s @ 5.9 keV \\ \hline
Cal B & Polarized, 15 $\times$ 15 mm$^2$ & $>$ 2.5 c/s @ 4.5 keV \\ \hline
Cal C & Unpolarized, diameter $<$ 5 mm & $>$ 23.1 c/s @ 5.9 keV \\ \hline
Cal D & Unpolarized, 15 $\times$ 15 mm$^2$ & $>$ 23.1 c/s @ 5.9 keV \\ \hline
Cal E & Unpolarized, 15 $\times$ 15 mm$^2$ & $>$ 23.1 c/s @ 1.7 keV \\ \hline
\end{tabular}
\end{table}

\begin{figure}[H]
\centering
\includegraphics[width=0.8\columnwidth]{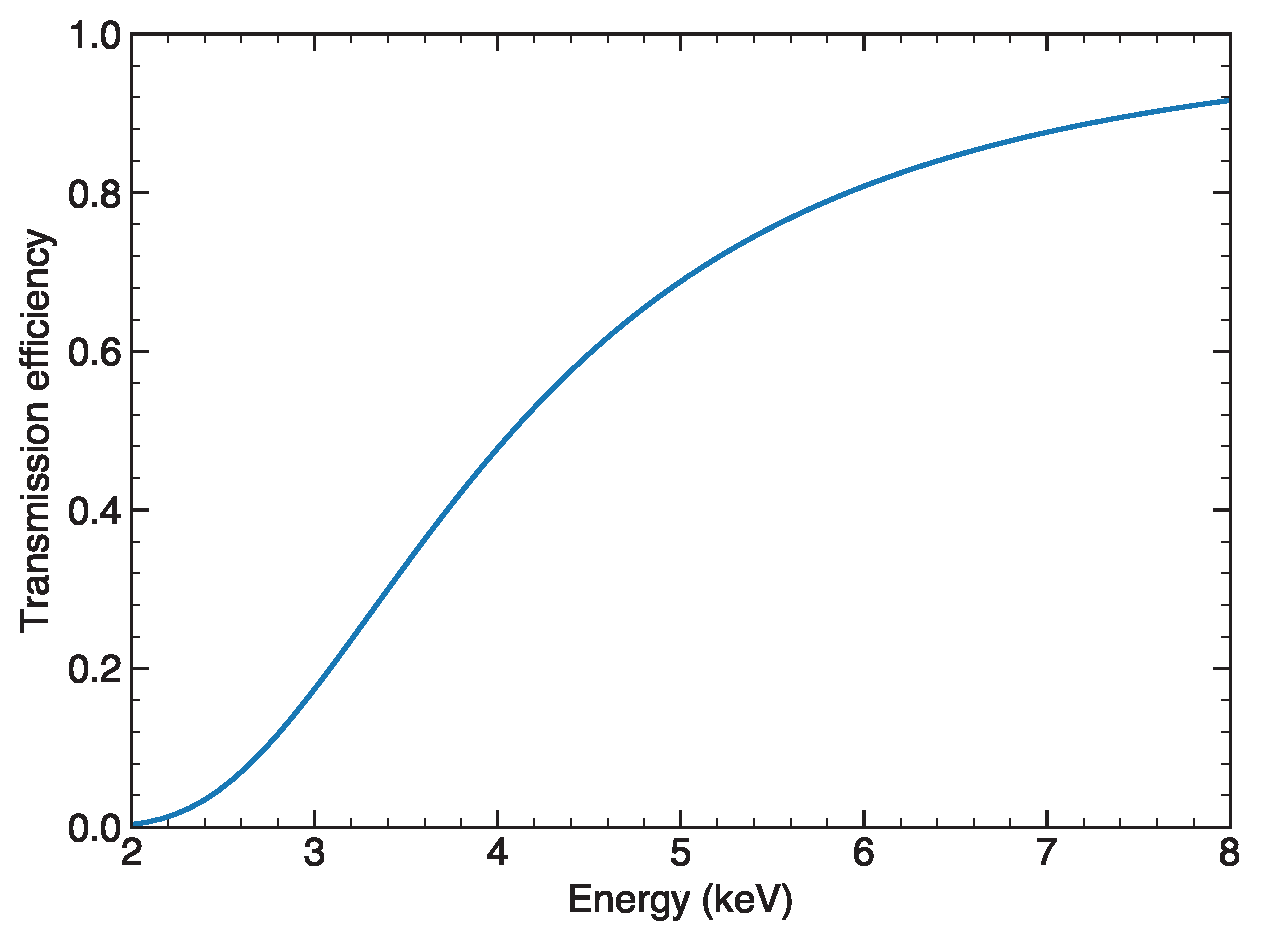}
\caption{Transmission efficiency of the gray filter designed for PFA.}
\label{fig:pfa_filter}
\end{figure}

The gray filter is intended for observations of very bright sources that may produce high count rates and lead to severe pile-up. The preliminary design consists of a Kapton foil with a thickness of 100~$\mu$m, coated with 100~nm aluminum on both sides. Its transmission is shown in Figure~\ref{fig:pfa_filter}. The filter will attenuate low-energy photons, with a transmission of about 40\% at 4 keV.

\subsubsection{Detectors and electronics.}
\label{sec:pfa:det}

\begin{figure}[H]
\centering
\includegraphics[width=\columnwidth]{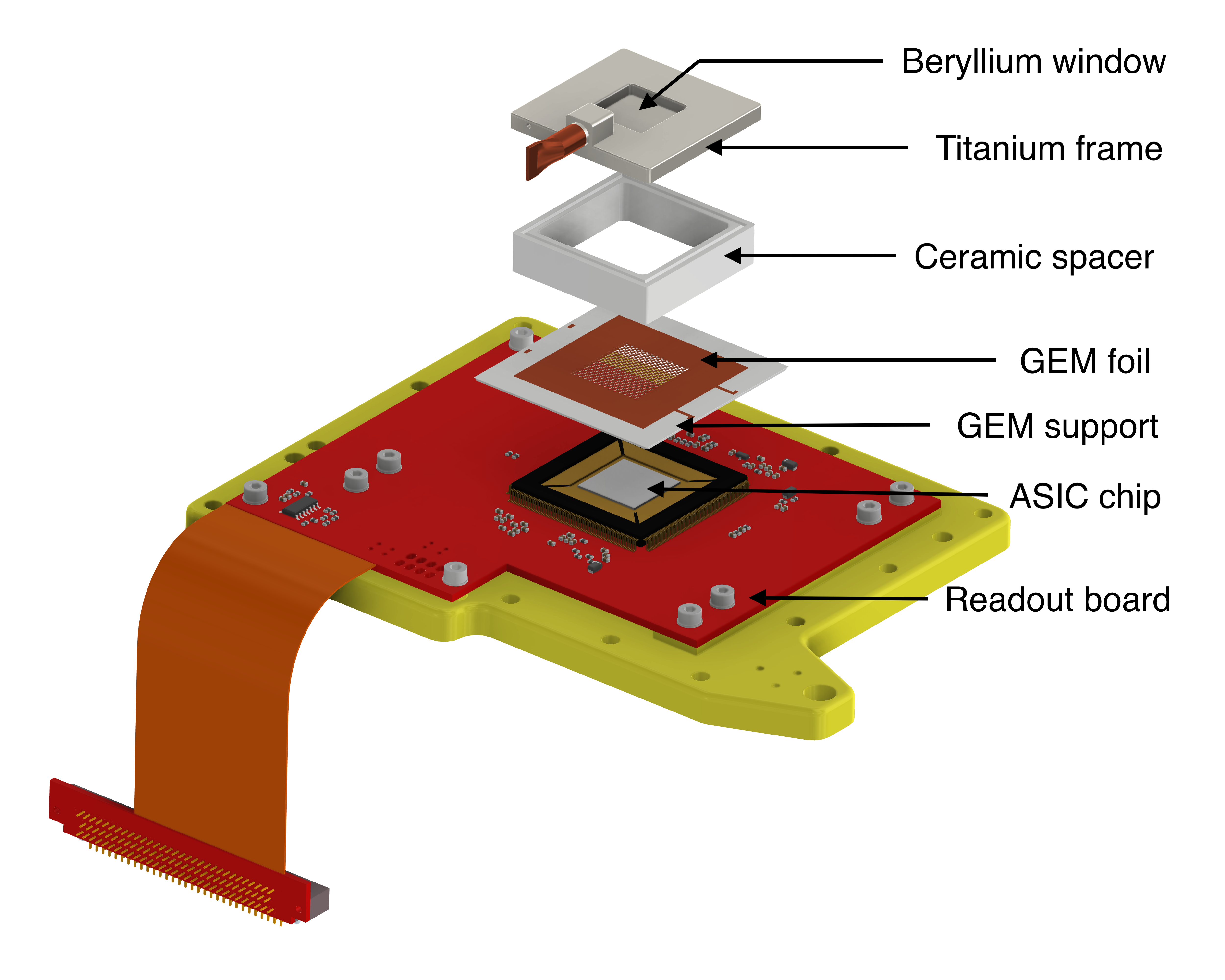}
\caption{A schematic drawing of the GPD.}
\label{fig:gpd}
\end{figure}

A GPD is adopted as the focal plane polarimeter for the PFA. It was originally developed by the INFN-Pisa group \cite{Costa2001,Bellazzini2003a,Bellazzini2007b,Bellazzini2013}, and has been used in PolarLight \cite{feng2019, li2021} and IXPE \cite{soffitta2021}. With the GPD, one can measure the 2D ionization track of the photoelectron in the gas chamber and infer the polarization of the X-ray source by reconstructing the emission angle of the photoelectron. The main characteristics of the GPD designed for PFA are summarized in Table~\ref{tab:GPD}, and a schematic drawing of the detector is shown in Figure~\ref{fig:gpd}.
\begin{table}[H]
\centering
\caption{Basic characteristics of the GPD for PFA.}
\footnotesize
\label{tab:GPD}
\begin{tabular}{l|l}
\bottomrule
\hline
\textbf{Parameter} & \textbf{Value} \\ \hline
Thickness of the Beryllium window & 50 $\mu$m \\ \hline
Thickness of the absorption gas & 10 mm \\ \hline
Thickness of the induction region & 0.8 mm \\ \hline
Active area & 15 $\times$ 15 mm$^2$ \\ \hline
Readout pitch & 50 $\mu$m (hexagonal) \\ \hline
Gas mixture & Pure DME \\ \hline
Filling pressure & 0.8 atm\\ \hline
Drift field strength & 2 kV/cm \\ \hline
Induction field strength & 5 kV/cm \\ \hline
Operation temperature & (20$\sim$25) $\pm$ 1 $^\circ$C\\ \hline
\end{tabular}
\end{table}

As shown in Figure~\ref{fig:gpd}, the GPD is a gas chamber sealed by a 50~$\mu$m thick Beryllium entrance window. The working gas is pure dimethyl ether (DME), which absorbs the incident X-rays and converts them to photoelectrons. A drift field of about 2~kV~cm$^{-1}$ is applied in the chamber to drive the secondary electrons ionized by the photoelectron to move towards the anode. 
To enable measurements with a sufficient signal-to-noise ratio, a gas electron multiplier (GEM) is mounted above the anode to multiply the number of electrons by a gain factor of a few hundred, which can be adjusted by the high voltage across the top and bottom layers of the GEM. The readout plane is mounted underneath the GEM. It is an ASIC chip~\cite{Bellazzini2004, Bellazzini2006b, Minuti2023} pixelated with a pitch of 50~$\mu$m, responsible for collection and measurement of the charges after multiplication. The ASIC has a dimension 1.5 cm $\times$ 1.5 cm, which defines the sensitive region of the detector. The readout noise is around 50~e$^-$. 
The BEE is designed to control and operate the ASIC, drive the analog to digital conversion, organize and store the data, and communicate with the satellite. They also regulate the high-voltage modules, which are needed for the drift field and to power the GEM field.

\begin{figure}[H]
\centering
\includegraphics[width=0.9\columnwidth]{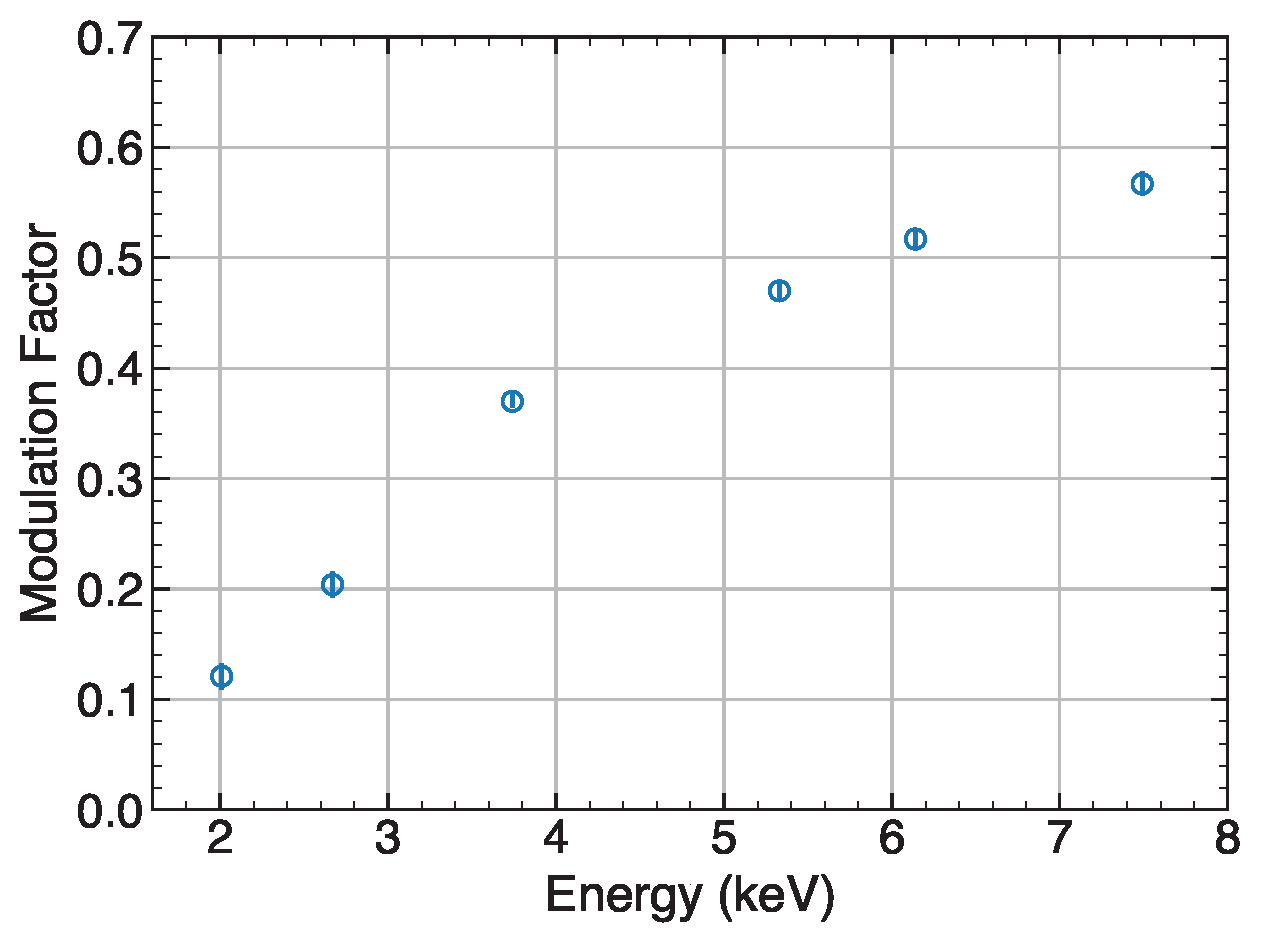}
\caption{Measured modulation factors of the GPD.}
\label{fig:pfa_mod}
\end{figure}

The modulation factor, defined as the instrument's response to a fully polarized X-ray source, is one of the key parameters of an X-ray polarimeter. Figure~\ref{fig:pfa_mod} shows the measured modulation factors of the GPD across the 2--8~keV energy range.
  
It decreases with decreasing energy for two reasons. First, the Coulomb scattering of the photoelectron by the nucleus will randomize the emission angle and lower the modulation degree, which becomes more important at low energies. Second, because of the shorter range of the track at low energies, the reconstruction of the emission angle becomes more difficult and is limited by the finite pixel size.

The GPD was first launched into orbit with the PolarLight project and operated for 3 years \cite{feng2019, li2021}. Then, it was used for the IXPE mission \cite{soffitta2021}, which has been working in orbit for more than 3.5 years as of May 2025. 
One of the issues with the GPD is the spurious modulation, an instrument systematics that shows polarization-like modulations when observing a non-polarized source. 
This systematics can be calibrated out before launch.
For IXPE, a dither pattern is employed during observation such that the focal spot covers a large area to further smooth out the systematics at different GPD regions. 
This approach can help reduce the calibration time. 
However, as eXTP has co-aligned SFA-T telescopes that have non-imaging SDD detectors, dithered observations are not preferred, unless the flux modulation caused by dithering in SFA-T can be modeled out effectively with the help of the co-aligned SFA-I and PFA with imaging capability. This issue remains to be studied during the phase C development.
Therefore, the calibration task for the PFA GPD is particularly challenging. The final calibration plan will be optimized on the basis of the actual spurious modulations of the GPD qualification model.

\subsubsection{Sensitivity}
\label{sec:pfa:perf}

\begin{figure}[H]
\centering
\includegraphics[width=\columnwidth]{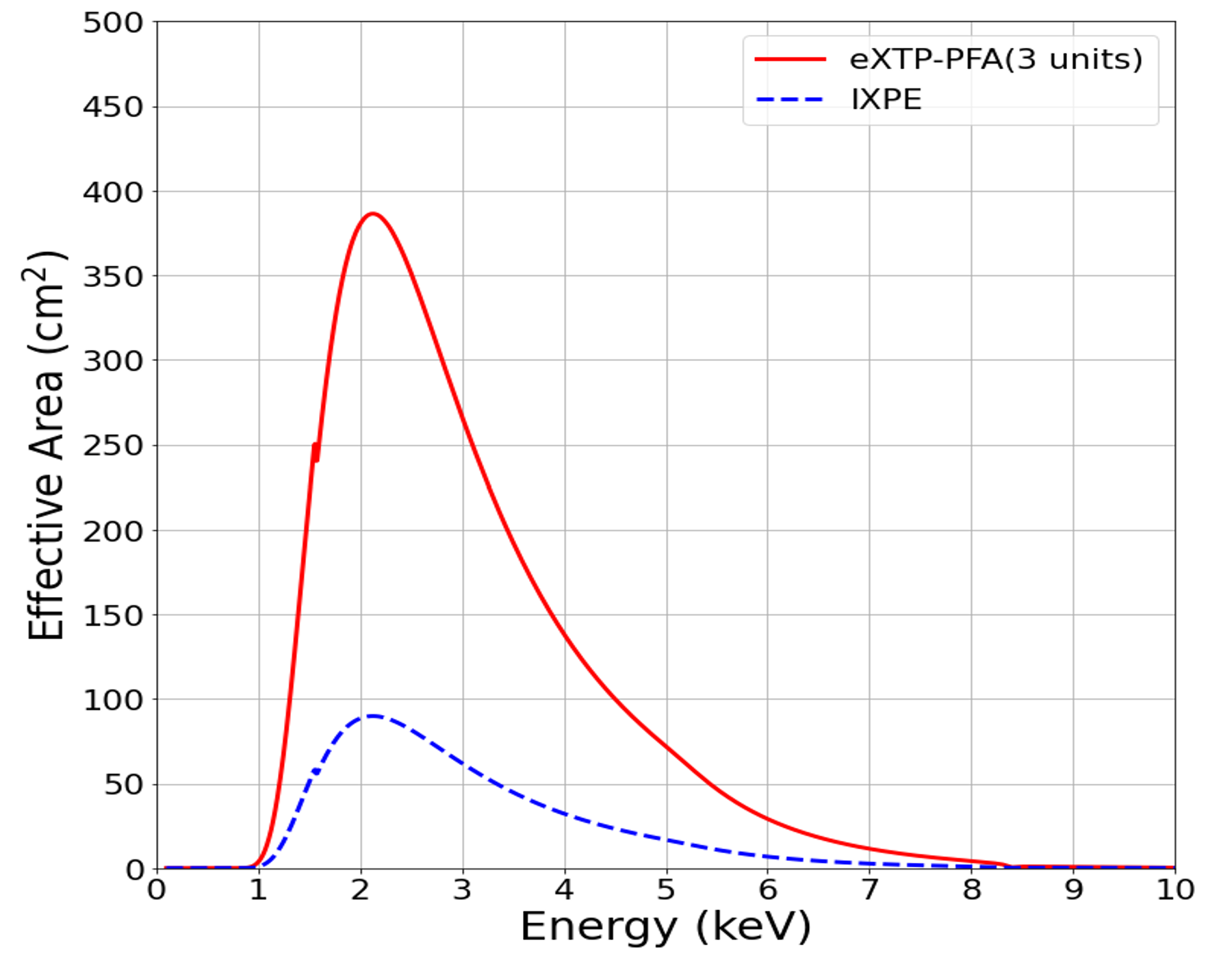}
\caption{Total effective area of PFA and IXPE.}
\label{fig:pha_eff_area}
\end{figure}

The total effective area of the three PFA telescopes as a function of energy is shown in Figure~\ref{fig:pha_eff_area}, against the effective area of IXPE for comparison. 
The effective area of the PFA is about five times that of IXPE at 3~keV. 

The sensitivity of the polarimeter is usually defined as the minimum detectable polarization (MDP), described as
$\textmd{MDP} = \frac{4.29}{\mu\,SA 
} \sqrt{\frac{SA + B}{t}}$, where $\mu$ is the modulation factor, $A$ is the effective area, $S$ is the source intensity in photons~cm$^{-2}$~s$^{-1}$, $B$ is the background rate in the source aperture in counts~s$^{-1}$, $t$ is the exposure time, and 4.29 corresponds to a confidence level of 99\%.  In cases where the background is not important,  the MDP can be approximated as $\textmd{MDP} = \frac{4.29}{\mu \sqrt{SAt}}$.
Thus, sensitivity is not only a function of the effective area, but is correlated with $\mu \sqrt{A}$. As one can see, the effective area peaks at about 2~keV while the sensitivity peaks at around 3~keV due to the increase of the modulation factor with energy. 


The sensitivity of PFA can be approximately estimated given a source with a crab-like spectrum. In the energy band of 2-8~keV, the mean modulation factor is $\sim$0.23 weighted by the observed spectrum. In the case of negligible background, an exposure of 1~ks of the Crab nebula will result in an MDP of 2\%. Sensitivities for observations with sources of different intensities and different exposure times can be scaled by simply using the equation for the MDP. We also note that the MDP quoted above indicates a detection at the confidence level 99\% (a chance probability of 1\% of having such a level of measurement from a fully unpolarized source), and more observation time is needed to achieve a precise measurement with a significance of 3$\sigma$ or more~\cite{Strohmayer2013}.

\subsection{Wide-field and Wide-band Camera -- W2C}\label{sec:W2C}

\begin{figure}[H]
\centering
\includegraphics[width=\columnwidth]{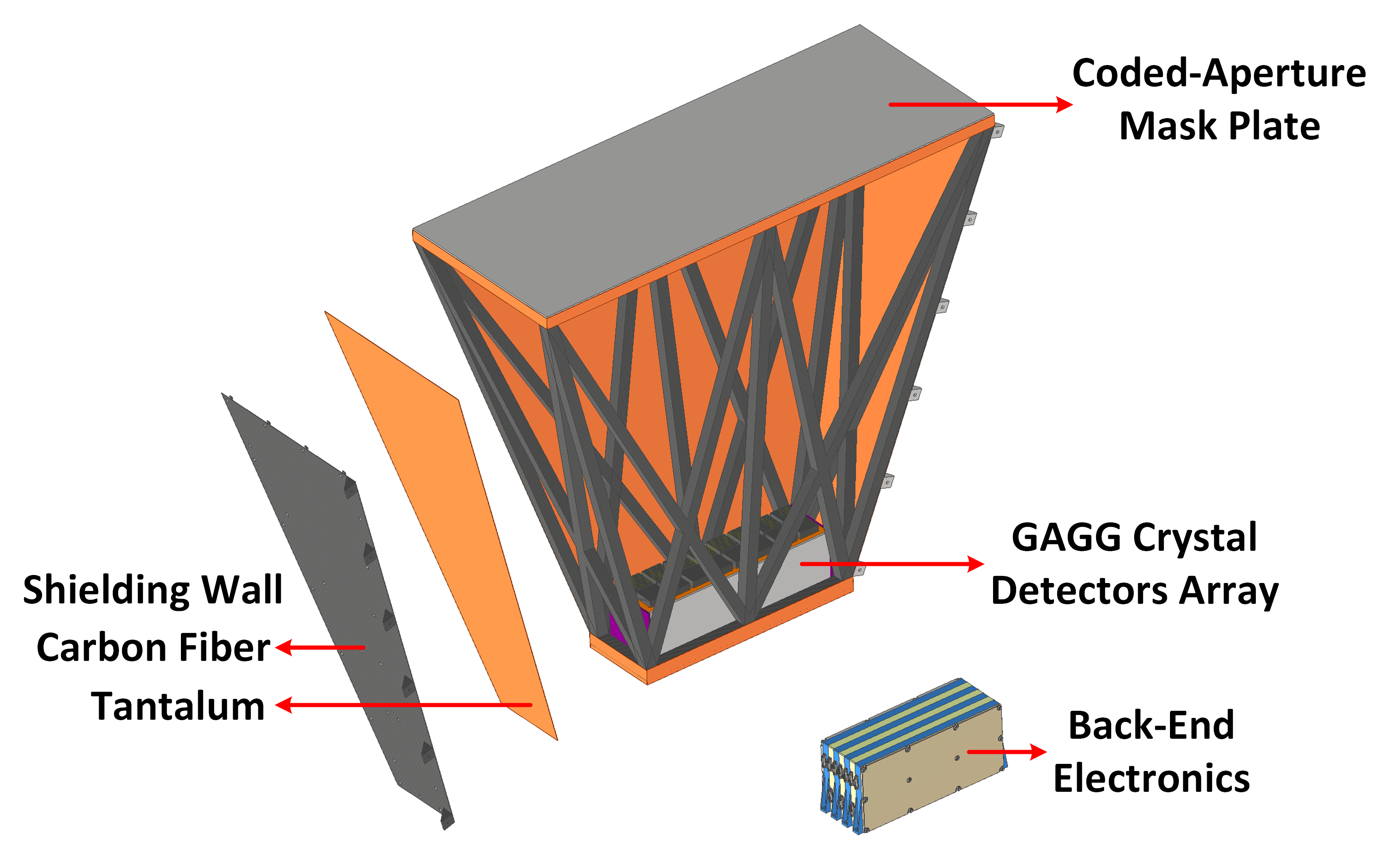}
\caption{Schematic drawing of W2C. W2C consists of four main parts: coded-aperture mask plate, GAGG detectors array, shielding walls and back-end electronics.}
\label{fig:w2c}
\end{figure}

The FoV of the W2C is approximately {1500} {square degrees} (Full-Width Zero Response, FWZR), with an energy range of {30}--{600} {keV}; the main technical specifications are listed in Table~\ref{tab:W2C}. Its primary function is to discover new high-energy bursts and trigger rapid onboard autonomous follow-up observations of the SFA and PFA. As shown in Figure~\ref{Fig:eXTP_satellite}, W2C is located at the bottom of the satellite, with its FoV not overlapping with that of SFA and PFA. The schematic drawing of the
W2C is shown in Figure~\ref{fig:w2c} for illustration. When detecting and locating a transient source, it will send trigger requests to the satellite platform. If approved, the satellite will rapidly reorient to position the transient source at the center of the SFA and the PFA's FoV. The main features of W2C include:

\begin{itemize}
\item Coded-aperture imaging and localization. It uses random coding patterns combined with partial coding FoV technology to expand the FoV of the image while maintaining a high signal-to-noise ratio in fully coded regions. It implements background subtraction and inverse matrix calculation of hot spots based on coding characteristics.  Because the mask blocking and detector interaction depths are both energy dependent, depth correction is also applied to both the coded mask and the detectors to improve localization accuracy, using the accurate energy spectrum of the detected gamma-ray burst spectral features to improve localization accuracy.

\item ASIC-controlled Silicon Photo-Multiplier (SiPM) readout. It employs high-light-output Gadolinium Aluminum Gallium Garnet doped with Cerium (GAGG) crystal arrays coupled with high quantum efficiency SiPM arrays, optimized for balanced detection efficiency in medium-high-energy gamma rays and low-energy X-ray fluorescence. It also features ASIC-based pixel-by-pixel readout of SiPM arrays with low electronic noise, ensuring 10--1000\,keV energy coverage. The readout electronics system originates from the POLAR-2 project \cite{Kole2025}.

\end{itemize}

\begin{table}[H]
\centering
\caption{W2C specifications}
\footnotesize
\label{tab:W2C}
\begin{tabular}{l|l}
\bottomrule
\hline
Field of View & Full coding: {49}{$^{\circ}$} $\times$ {9.6}{$^{\circ}$} \\
 & Half coding: {60}{$^{\circ}$} $\times$ {16}{$^{\circ}$} \\
 & FWZR: {68}{$^{\circ}$} $\times$ {22}{$^{\circ}$} \\ \hline
Angular Resolution & $\sim${20}{$^\prime$}  @ 30-100 keV\\ \hline
Positioning Accuracy & $\sim${5}{$^\prime$} @ 30-100 keV\\ \hline
Effective Area & $\sim${160}{cm$^{2}$} @ {60}{keV} (normal incidence) \\ \hline
Sensitivity & $\sim4\times10^{-7}$ {erg$\cdot$ cm$^{-2}\cdot$ s$^{-1}$} @ {10}--{1000} {keV} in {1} {s} \\ \hline
Energy Range & {30}--{600} {keV} \\ \hline
Energy Resolution & $\le${30}{\%} @ {60}{keV} \\ \hline
Time Resolution & $\le${25} $\mu$s \\ \hline
\end{tabular}
\end{table}

\begin{figure}[H]
\centering
\includegraphics[width=\columnwidth]{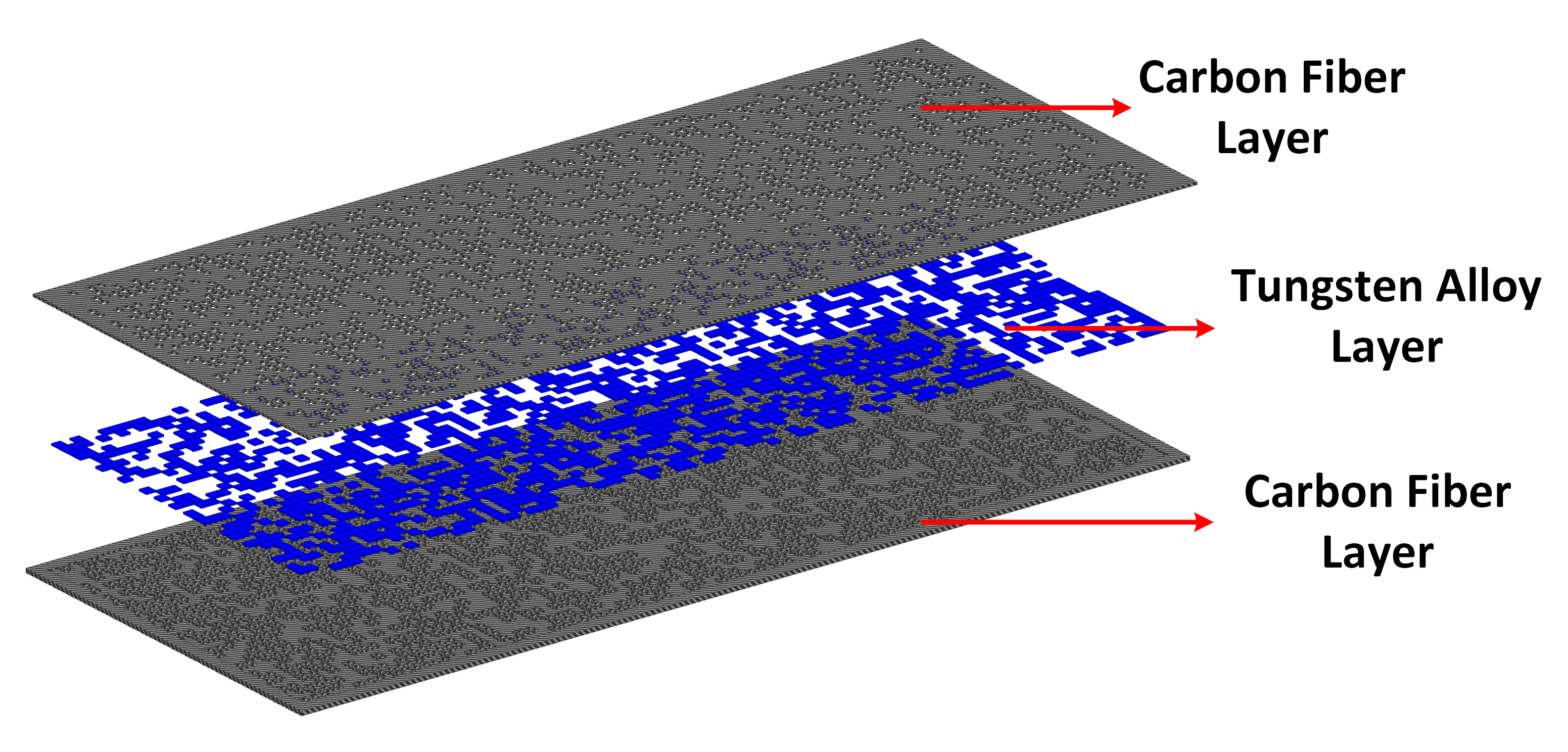}
\caption{Design of the the coded-aperture mask plate of W2C.}
\label{fig:w2c_mask}
\end{figure}

The coded-aperture mask plate of W2C is primarily composed of three layers of materials: the top layer is a 1.5~mm thick carbon fiber, the middle layer consists of randomly distributed 1~mm thick tungsten alloy block pieces, and the bottom layer is another 2.5~mm thick carbon fiber material. The upper and lower layers of carbon fiber have randomly placed apertures with an open area ratio of $50\%$. X-rays passing through apertures within the W2C field of view will generate hits on the detector, while X-rays blocked by tungsten alloy blocks in the middle layer within a certain energy range ($\le$100 keV) will not be detected. Note that the mask does not block X-ray photons above 100 keV effectively, in order to minimize the weight of the mask and maximize the spectral measurement capability to transients at energies above 100 keV. The size of the tungsten alloy blocks and the apertures matches the GAGG pixel size, both being 5.75 mm × 5.75 mm. The design of the W2C coded-aperture mask plate is shown in Figure~\ref{fig:w2c_mask}.

In addition to the coded-aperture mask plate at the top, the W2C payload is surrounded by shielding walls on all sides to block most of the X-ray photons and low-energy charged particles in orbit, thereby ensuring the sensitivity and imaging localization accuracy. As shown in Figure~\ref{fig:w2c}, the wall of shielding in W2C consists mainly of two layers: the outer layer is made of 1~mm thick carbon fiber material, providing structural support, and the inner layer is made of 1~mm thick tantalum, serving as the primary shielding material.

\begin{figure}[H]
\centering
\includegraphics[width=0.8\columnwidth]{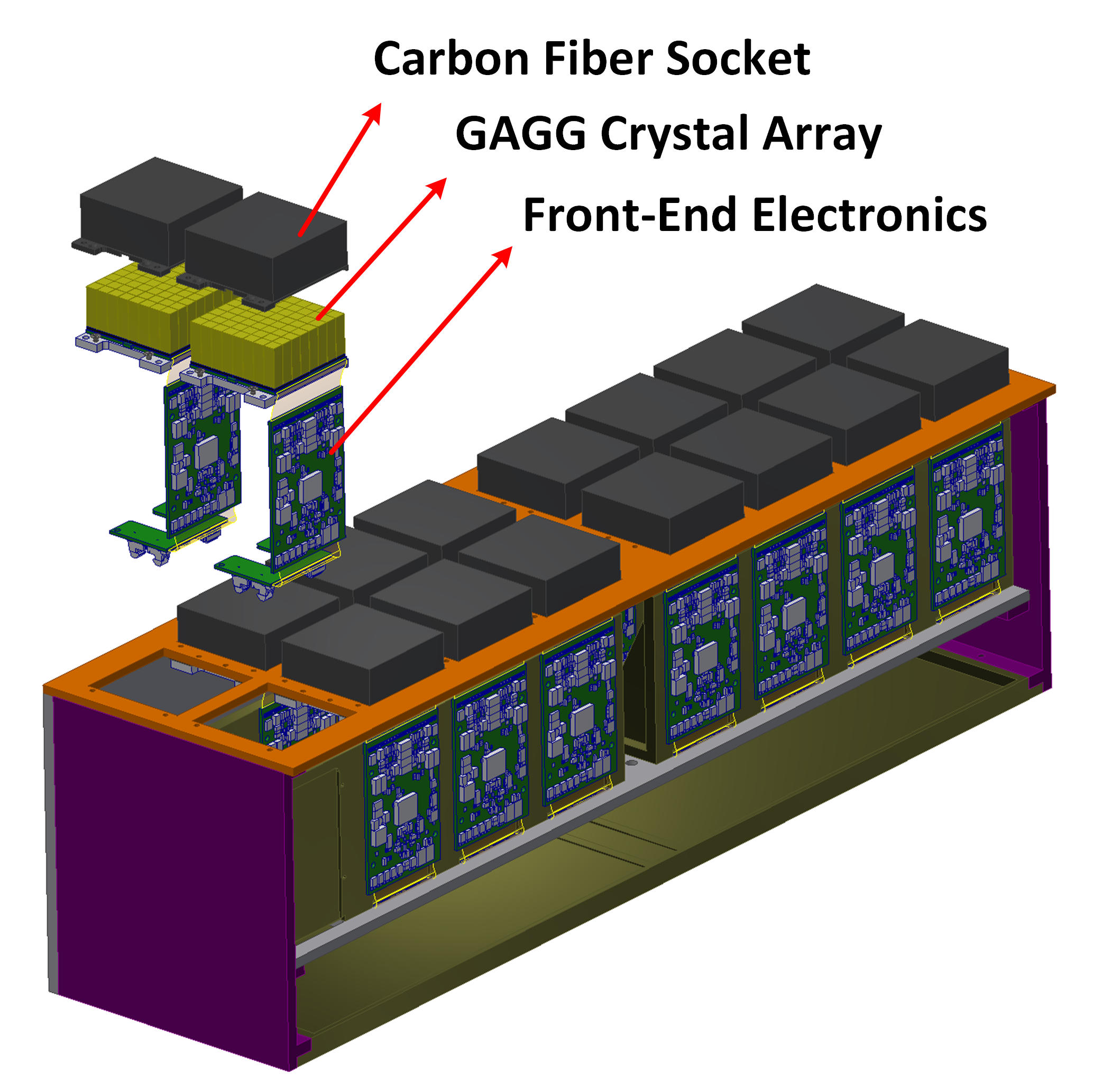}
\caption{Schematic diagram of the GAGG crystal detector array of W2C.}
\label{fig:w2c_gagg_detector}
\end{figure}

The W2C GAGG scintillator detector array consists of a total of 16 detector units organized in a 2×8 array. This configuration is primarily designed to accommodate the spatial constraints of the satellite platform. Each detector unit is composed of 64 GAGG scintillator crystal pixels in an 8×8 array. The dimension of each pixel is 5.75~mm×5.75~mm×20~mm, and Enhanced Specular Reflector (ESR) reflective film is used between the pixels to reflect and isolate the fluorescent signals generated within the crystals. The pitch between adjacent pixels is 6.25~mm. The GAGG crystal array module is inserted in a carbon fiber socket for mechanical support and installation purposes. The signals generated within the GAGG crystals are collected and read out by the SiPM array board of the front-end electronics. The SiPM array board consists mainly of 64 small pieces of SiPMs from Hamamatsu, with each SiPM corresponding to the signal readout of one GAGG pixel. In addition to the SiPM board, the front-end electronics also includes an ASIC board and an interface board. The ASIC board contains key electronic components, such as the ADC and FPGA for reading out the signals from the SiPMs, while the interface board is used for communication with the back-end electronics. A schematic diagram of the W2C GAGG crystal detector array is shown in Figure~\ref{fig:w2c_gagg_detector}.

The back-end electronics system of W2C is primarily used for payload triggering and the implementation of imaging algorithms, as well as transmitting payload data to the satellite platform, providing control commands, and managing power supply for the payload. Its main functions include:
\begin{itemize}
\item Communicating with the 16 detector units to read the physical event data packets.

\item Managing science data, implementing trigger and localization functions, and forming event-by-event data packets.

\item Performing remote control, telemetry, and status configuration for the detector units.

\item Interfacing with the platform to process data, complete trigger, localize, and send science data packets.

\item Interfacing with the platform computer for remote control and telemetry, completing the sending and receiving of remote control and telemetry data packets.

\item Receiving the PPS signal from the platform computer to achieve system time synchronization.

\item Implementing temperature telemetry and heater power supply functions for the temperature control module.

\item Managing the primary power distribution for the system.

\item Switching the system's operational modes.
\end{itemize}

\begin{figure}[H]
\centering
\includegraphics[width=\columnwidth]{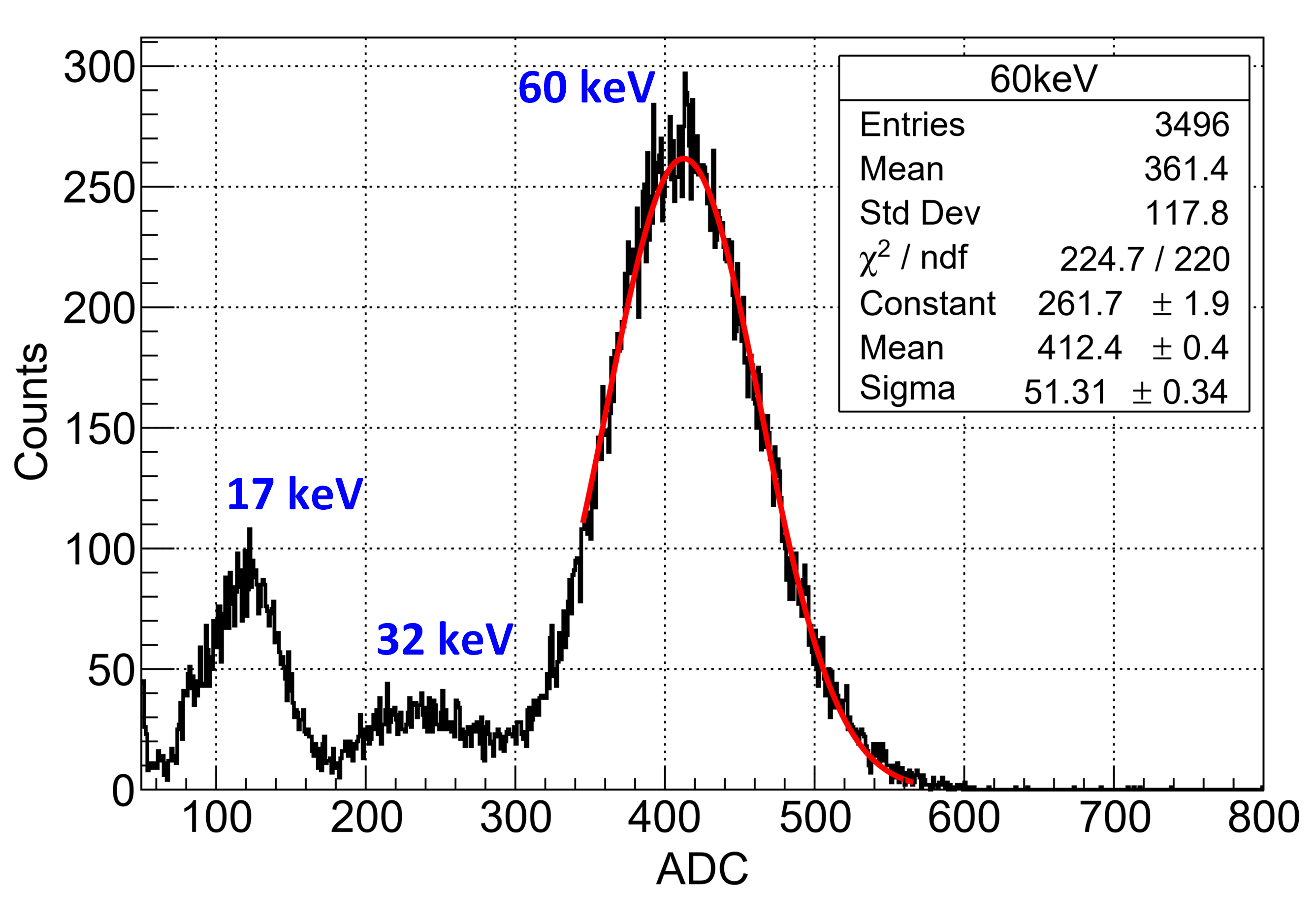}
\caption{Measured spectrum of the GAGG scintillator detector unit prototype with a hard X-ray beam. Black curve: measured data. Red curve: fitted with Gaussian function. The beam energy is 60~keV. Three peaks are visible in the spectrum: the 60~keV peak, the 32~keV fluorescence from Barium due to the $BaSO_{4}$ housing, and the 17~keV escape peak.}
\label{fig:w2c_gagg_spectrum_60keV}
\end{figure}

Using a prototype of the GAGG scintillator detector unit (with GAGG crystal pixel dimensions of 5.75~mm × 5.75~mm × 5~mm), preliminary performance tests were conducted with a hard X-ray beam. Figure~\ref{fig:w2c_gagg_spectrum_60keV} shows the spectrum measured with this detector unit obtained using a 60~keV hard X-ray beam. In addition to the 60~keV beam energy peak in the spectrum, there is also an escape peak of approximately 17~keV due to the $K_\alpha$ characteristic X-ray energy of Gadolinium (Gd) in GAGG, which is about 43~keV. Furthermore, since barium sulfate (BaSO$_{4}$) was used as the reflective layer material between the GAGG crystal pixels in this detector unit prototype, the measured spectrum also includes a barium fluorescence line that corresponds to about 32~keV.

The large FoV of W2C provides unique opportunities to detect $\sim15$ GRBs per year. The burst alert system is modelled following the design of the SVOM mission \cite{Atteia2022} and elements of the INTEGRAL BAS \cite{Mereghetti2003}. 

Two onboard trigger algorithms are foreseen: an image trigger, which performs systematic image deconvolution on long time scales, and a count rate trigger, which as a first step selects short time scales of counts in excess over background to be deconvolved in a second step. The detection of an uncatalogued source gives rise to a localized burst candidate. 

However, W2C has not yet been adopted as a primary eXTP payload, due to some programmatic issues; it will likely be approved in a later stage and eventually installed onboard the eXTP platform as a secondary instrument. Currently, the required weight, power, and interfaces have been reserved in the design of the satellite platform. It is also possible to double the scale of W2C and mount it on the top surface of the platform (its FoV is 
co-aligned with that of SFA and PFA), if a different and more powerful launcher is used to replace the current baseline launcher CZ-3B G5 described in Section~\ref{subsec:laucher}.

\section{The mission profile}\label{sec:Mission}
\subsection{Satellite System}
\begin{figure}[H]
\centering
\includegraphics[width=0.85\columnwidth]{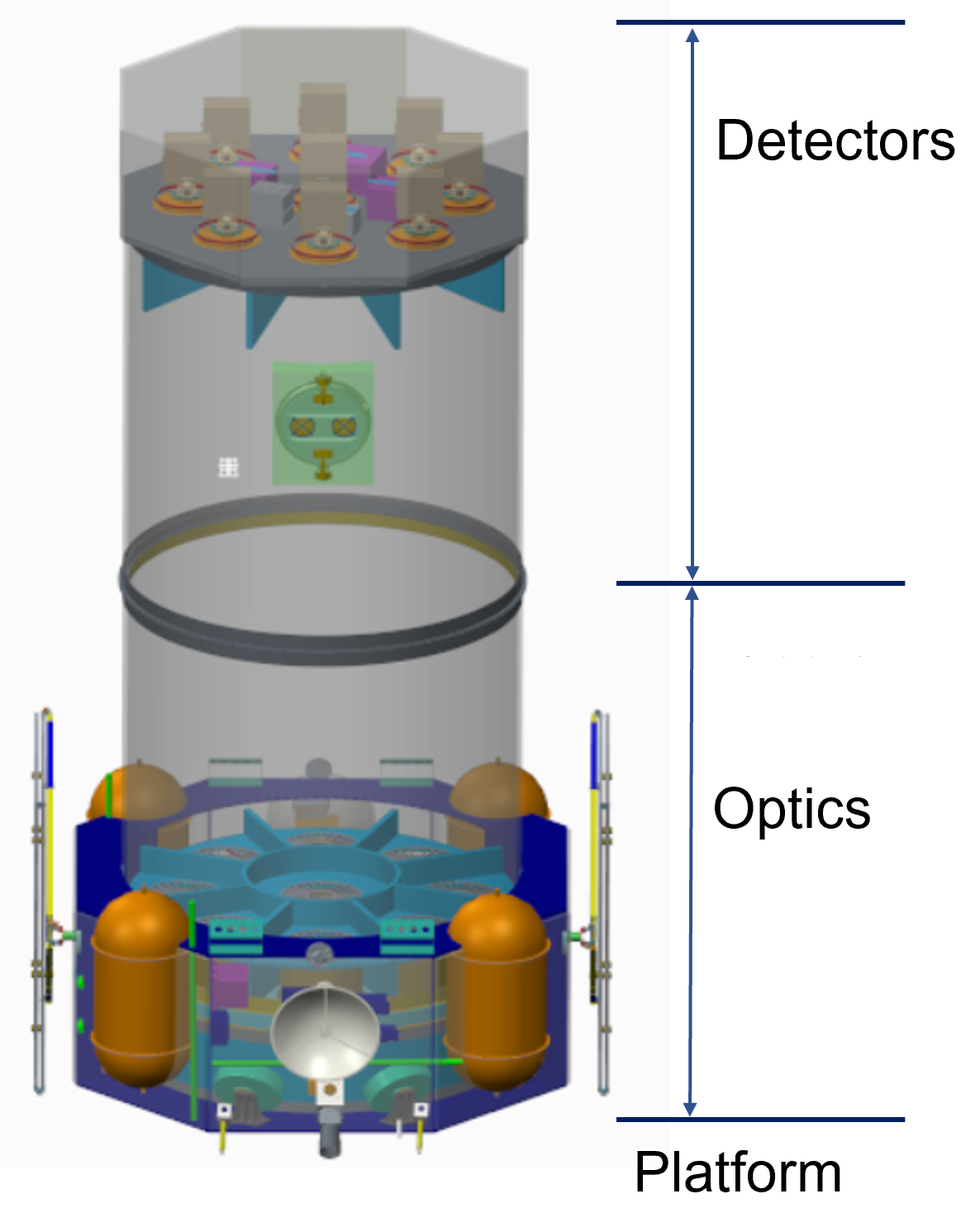}
\caption{The eXTP satellite stowed during launch (diameter 3755 mm x 6604 mm).}
\label{fig:eXTP_stowed}
\end{figure}

\begin{figure}[H]
\centering
\includegraphics[width=0.85\columnwidth]{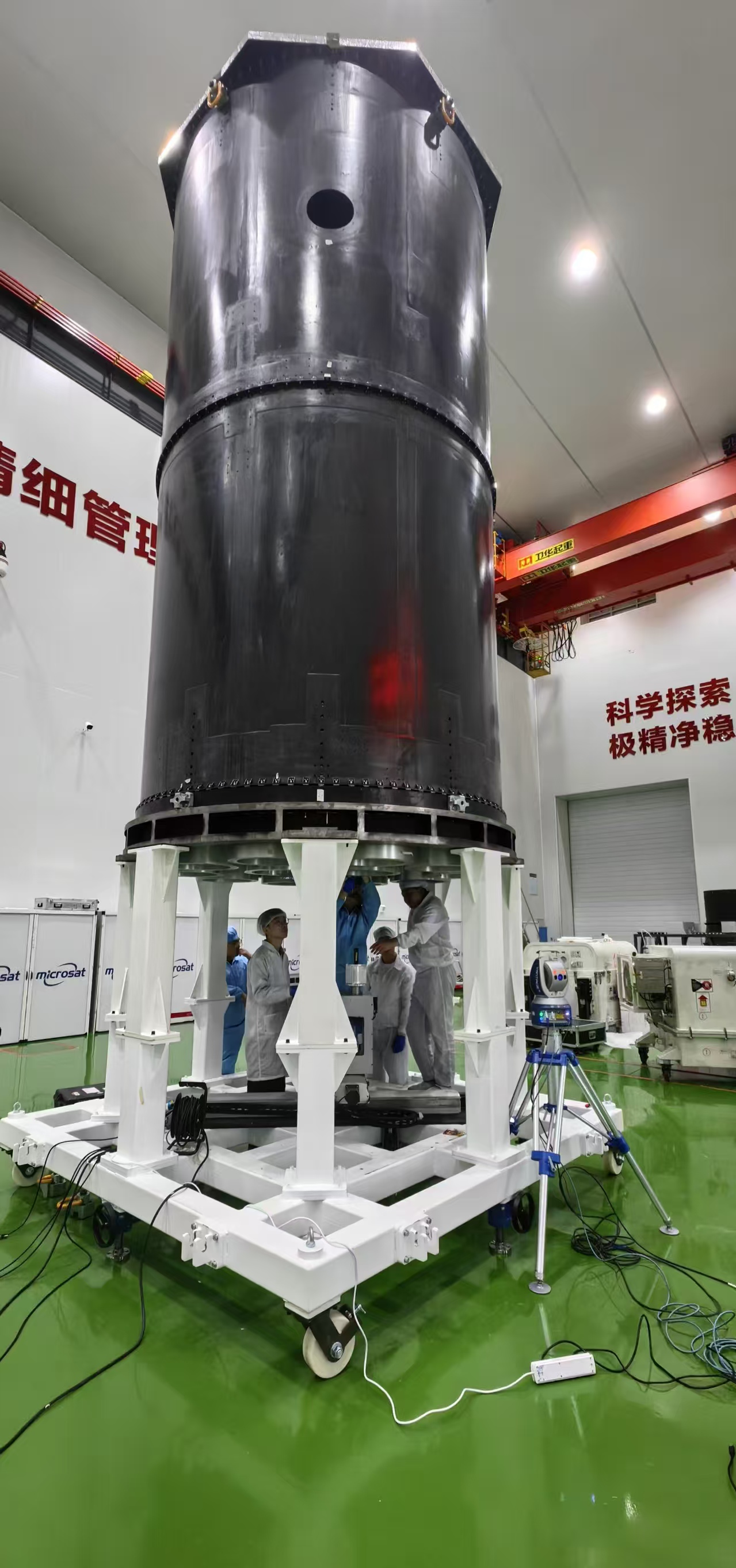}
\caption{The main structure of the eXTP satellite, in which the mirrors and focal plane cameras are accommodated. Note that the structural models of mirrors (just above the engineers) and cameras (in the top) were installed for mechanical vibration tests. The tests results show that the mirrors and cameras were still well aligned within the requirements after vibrations.}
\label{fig:main-structure}
\end{figure}

\begin{figure}[H]
\centering
\includegraphics[width=1.0\columnwidth]{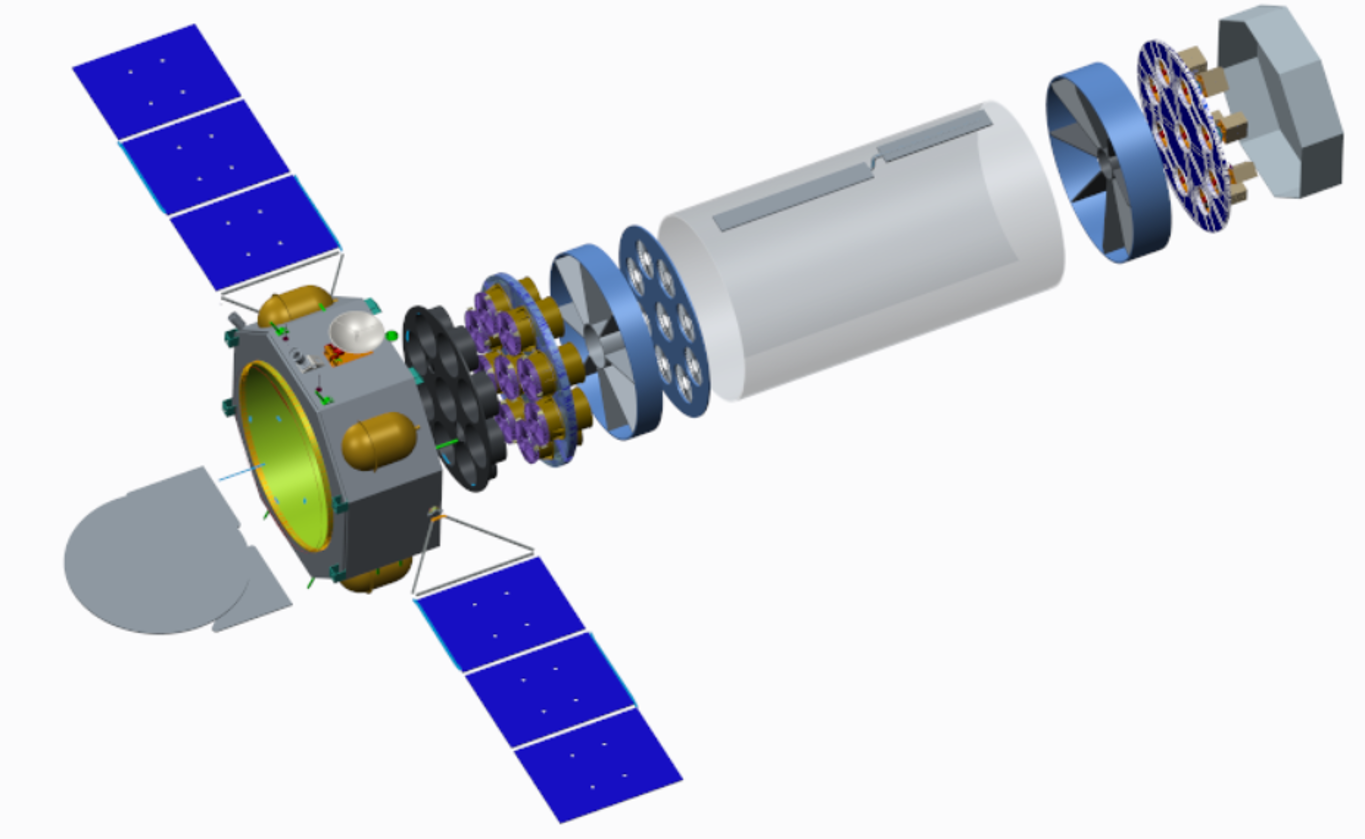}
\caption{The expanded view of the deployed satellite is shown. In orbit the eXTP encumbrance is 4255 mm x 8800.9 mm x 12958.3 mm.}
\label{Fig:eXTP_expanded}
\end{figure}

The eXTP satellite stowed during launch is shown in Figure~\ref{fig:eXTP_stowed}. Figure~\ref{fig:main-structure} is a picture of the main structure of the eXTP satellite, in which the mirrors and focal plane cameras are accommodated. The deployed configuration is shown in Figure \ref{Fig:eXTP_expanded}.

The main specifications of the satellite system are as follows:
\begin{itemize}
\item Orbit Altitude: Apogee: $\ge$100,000 km, Perigee: $\sim$5,000 km
\item Orbit Type: High Elliptical Orbit
\item Orbit Inclination: $\sim28.5^{\circ}$
\item Total Satellite Mas: $\le$3,970 kg 
\item Focusing Mirror Pointing (3$\sigma$): $\le$1.1$^\prime$ (SFA\&PFA, deviation of each focusing mirror's optical axis relative to celestial target)
\item Focal Plane Detector Position (3$\sigma$): Axial $\le$1 mm, Radial $\le$0.5 mm (relative to focusing mirror's focal point)
\item Pointing Method: Three-Axis Stabilized Inertial Pointing 
\item Pointing Accuracy: $\le$3$^{\prime\prime}$  (3$\sigma$)
\item Attitude Stability: $\le$2$^{\prime\prime}$/s (3$\sigma$)
\item Attitude Measurement Accuracy: $\le$1$^{\prime\prime}$ (3$\sigma$) (Star Sensor)
\item Micro-Vibration: $\le$1.2$^{\prime\prime}$ (rms, 10-2000 Hz)
\item Manoeuvring Capability: 30$^{\circ}$ with stabilization in 10 minutes
\item Transmission Band: Ku-band Earth Communication
\item Data Rate: 560/1120 Mbps (Ku) 
\item Storage Capacity: Supports 2-day scientific data storage 
\item Control System: USB Telemetry
\item Uplink Rate: 2000 bps
\item Downlink Rate: 16384/8192/4096/2048 bps
\item Particulate Contamination: $\le$1000 ppm (Target: 300 ppm)
\item Molecular Contamination: Focusing Mirror $\le 1\times 10^{-7}$ g/cm$^2$  Detector $\le 1\times 10^{-6}$ g/cm$^2$
\item Onboard Time Accuracy (GNSS available): $\le$1 $\mu$s
\item Processing Speed: Processor frequency $\ge$48 MHz
\item Design Lifetime:  5-year nominal, additional 3-year extended
\end{itemize}

\subsection{Launch Vehicle System}\label{subsec:laucher}
A CZ-3B G5 or equivalent launch vehicle will be used to launch the eXTP satellite into a 200$\times$100,000 km initial orbit. The technical requirements are as follows:
\begin{itemize}
    \item {Payload capacity:} $\ge$3970 kg (excluding interface brackets, adapters, and transition cables)
    \item {Satellite structural stiffness:} Lateral frequency (1st order): $>10$\,Hz; Longitudinal frequency (1st order): $>30$\,Hz
    \item {Fairing requirements:} Net envelope: $\ge$3850 mm (transport with fairing from technical area to launch pad)
    \item Environmental conditions:  Temperature: 18–25$^{\circ}$C; Humidity: 30\%–45\%; Cleanliness: Class 10,000
    \item Thermal/pressure requirements: Radiation heat flux $\le$700 W/m$^2$ during ascent; Free molecular heating $\le$1135 W/m$^2$ during fairing separation; Maximum pressure drop rate $\le$6.9 kPa/s
    \item {Orbit insertion accuracy:} Semi-major axis deviation: $\le$800 km; Perigee altitude deviation: $\le$41 km; Inclination deviation: $\le$0.22$^{\circ}$; Argument of perigee deviation: $\le$0.71$^{\circ}$
RAAN deviation: $\le$1.09$^{\circ}$
    \item {Separation requirements:} Attitude: +X axis earth-facing, +Y axis along velocity vector; Relative velocity: $\ge$0.5 m/s; Attitude angle error (3$\sigma$): $\le$2$^{\circ}$ (all axes); Angular rate error (3$\sigma$): $\le$0.5\%/s (all axes)
\end{itemize}

\subsection{Launch Site System}
The launch site system is responsible for rocket/satellite testing, launch operations, and facility management. The technical requirements are as follows:
\begin{itemize}
    \item {Test complex:} Environment: Class 100,000 cleanliness, 18–25$^{\circ}$C, 30\%–45\% RH; Nitrogen supply: Support for satellite's high-purity nitrogen system
    \item {Launch area:} Fairing environment: Class 10,000 cleanliness, 20$\pm$3$^{\circ}$C, 30\%–50\% RH; Nitrogen interface: Dedicated purge system on launch tower
\end{itemize}

\subsection{Tracking \& Control System}
The technical specifications for the tracking \& control system are as follows:
\begin{itemize}
    \item {USB system:} Uplink rate: 2000 bps (BER $\le$1$\times$10$^{-6})$; Downlink rate: 16384 bps (BER $\le$1$\times$10$^{-5}$)
   \item {Initial tracking:} $\ge$110s contact at 5$^{\circ}$ elevation (TBC)
   \item {Routine operations:} $\ge$4 daily contacts ($\ge$30 minutes total); CCSDS-compliant telemetry/command formats
   \item {ToO response:} Routine ToO: $\le$12 hours implementation; Priority ToO: Next available contact; $\le$2 ToO commands/day
   \item {Orbit determination (3$\sigma$):} 1-orbit prediction error: $\le$3.5 km; 2-orbit prediction error: $\le$5 km; Post-processed accuracy: 1–3 km position, $\le$1 m/s velocity
   \item {Time calibration:} Correct when the clock drifts more than 500 ms (with 5 ms accuracy)
\end{itemize}

 Here ToO refers to external triggers received in ground. Autonomous pointing can be triggered by transients detected with W2C onboard. On the other hand, if a LEO is eventually chosen, ToO responses to external triggers may also be very fast, similar to the EP and SVOM missions.
\subsection{Ground Support System}
The primary tasks of the Ground Support System (GSS) are: (1) to organize and complete the planning and design of in-orbit operation schemes (including Ku-band data transmission station selection) based on scientific requirements and satellite payload operation modes; (2) complete the reception, recording, and ground transmission of downlinked data; (3) carry out scientific observation mission planning, schedule development, command generation (including Target of Opportunity (ToO) event handling), mission monitoring, and status analysis for comprehensive operations and management, ensuring in-orbit operation control of scientific missions; (4) process and analyse the quality of satellite downlinked data, produce edited-level products, and distribute them to the Scientific Application System; and (5) collect, archive, and manage scientific data products, provide long-term data release services, establish a scientific research support environment, and assist scientific users.

The key capabilities of the GSS are as follows:
\begin{itemize}
    \item 5-year primary + 3-year extended mission support
    \item Ku-band ground stations (QPSK modulation)
    \item $\le$4 hr routine/$\le$1 hr emergency command generation
    \item 10 min ToO command processing
    \item Data processing within 1 hr of acquisition
    \item $\ge$6 PB archival storage capacity
\end{itemize}

\subsection{eXTP Science Data Center}
The Science Data Center (SDC) is responsible for (1) soliciting and selecting observation proposals for the eXTP satellite, formulating scientific observation plans (including routine and ToO observation plans); (2) monitoring payload operations, conducting performance analysis, and coordinating investigations into anomalies; (3) developing eXTP quality models, performing ground-based simulations, establishing calibration databases using ground calibration results, and updating in-orbit calibration analyses and databases;(4)  processing scientific observation data to generate advanced scientific data products and auxiliary data products; (5) developing eXTP data analysis software and providing technical support for data analysis; (6) designing mission-level end-to-end scientific simulation systems to produce end-to-end simulated scientific data; and (7) organizing research and application of scientific observation data; managing the release and storage of scientific data products and mission information during the mission operation phase.  The eXTP data storage and release, as well as service to the community, will be the sole responsibility of the National Space Science Center (NSSC) of CAS after the mission operation is terminated; actually NSSC will also store all eXTP data and can also release the data during the mission operation phase.

To achieve the scientific objectives of the eXTP quickly after satellite launch, the SDC must also conduct pre-research during the development phase, including completing theoretical models, simulations, development of specialized analysis tools, and development of a preliminary observation plan tuned to reach the scientific goals of the eXTP. This involves (1) in-depth studies of the fundamental properties and radiative characteristics of compact celestial objects, refining existing theoretical models and research methods, and simulating scientific objectives; (2) developing dedicated physical analysis tools based on multi-method observational results (light curves, energy spectra, images, modulation curves, etc.); and (3) formulating preliminary observation requirements and curating scientific observation source catalogue.

The requirements for the SDC are as follows:
\begin{itemize}
    \item Mission planning for 5+3 year operations
    \item ToO response: $\le$13 hr implementation for routine ToO; $\le$20 min planning for priority ToO
   \item Payload monitoring: $\le$5 s anomaly alert
   \item Data processing: $\le$8 hr calibration product generation
End-to-end simulation system
   \item Products: $\ge$5 PB data storage; white papers
\end{itemize}

\section{Observing strategy}\label{sec:4}
The primary objective of the observation strategy is to ensure the best possible scientific outcome for the mission, opening the mission to the scientific community, while ensuring a return to the institutions and countries that financially supported the mission. Therefore, it is foreseen that eXTP will operate as an observatory open to the scientific community. Observation time will be allocated through annual announcement of opportunities and through scientific peer review. A fraction of the observing time will be allocated to key core projects that should ensure that the core science goals of the mission are reached. The mission will be designed to react fast to transient events, e.g., outbursts, performing flexible ToOs. ToOs (outside the state changes of sources defined in the guest observer program and in the key projects) can be proposed by the scientific community, and the Principal Investigator (PI) of eXTP will decide whether to schedule the proposed observation, after consulting the relevant key members of the eXTP team. Depending on the nature of the ToO, either the one-year proprietary data rule can be applied, or the data will be made available to the science community in general. A standard one-year proprietary data rights are assumed for the pointed instruments, and after this period all data shall become public. For the W2C no proprietary data rights will apply, and these data should be made available to the community on a much shorter time scale. During routine operations, several pre-products data will be produced typically in 3 hours after the observation. The data of the Burst Alert System will be distributed to the community as soon as possible using the appropriate systems, and no data rights will be applied.

\section{Conclusions}\label{sec:5}
In this paper we have presented the main scientific goals, technical and technological aspects of the scientific payload of the eXTP mission, the performance of the instruments, and the main aspects of the mission. The eXTP mission is designed to address key questions of physics in the extreme conditions of ultra-dense matter, strong field gravity, and the strongest magnetic field existing in nature. The payload includes two narrow field instruments: the SFA, characterized by a large area at soft energies, and the PFA, an X-ray polarimeter whose effective area is about five times that of the polarimeter onboard the IXPE mission \cite{IXPE}. The science payload is completed by wide field monitoring capabilities. This combination will enable for the first time ever spectral-timing-polarimetry studies of a large number of objects populating the time variable Universe. In addition to investigating fundamental physics, eXTP will play a major role in the time-domain and multi-messenger era and become a very powerful observatory for astrophysics, providing data on a variety of galactic and extragalactic objects. In particular, the W2C will be able to detect the electro-magnetic counterparts of transient sources of the multi-messenger Universe including gravitational wave sources and cosmic neutrinos sources.

The mission is led by China and is executed under the management of the CAS National Space Science Center, within the framework of the activities of the CAS Bureau of Major R\&D Programs. IHEP of CAS leads the science consortium and is responsible for coordinating
all science payload contributions within China and from international partner countries and agencies.


\emph{Acknowledgements.} The  Chinese  team  acknowledges  the  support of the National Natural Science Foundation of China (Grant No. 12333007), the International Partnership Program of Chinese Academy of Sciences (Grant No.113111KYSB20190020) and the Strategic Priority Research Program of the Chinese Academy of Sciences (Grant No. XDA15020100). The Italian collaboration acknowledges support by ASI, under the dedicated eXTP agreements and agreement ASI-INAF n.2017-14-H.O., by INAF and INFN under project REDSOX. The German team acknowledges support from the Deutsche Zentrum f{\"u}r Luft- und Raumfahrt, the German Aerospce Center (DLR). 
The Spanish authors acknowledge support from MINECO grant ESP2017-82674-R and FEDER funds.


\emph{Conflict of Interest}  The authors declare that they have no conflict of interest.












\end{multicols}
\end{document}